\documentclass[fleqn,usenatbib]{mnras}

\usepackage{amsmath,amssymb,amstext}
\usepackage{subfigure}
\usepackage{float}
\usepackage[version=4]{mhchem}
\usepackage{multirow}
\usepackage[utf8]{inputenc}
\usepackage{lmodern}
\usepackage[T1]{fontenc}
\usepackage{mathtools,nccmath}
\usepackage{environ}
\usepackage{times}
\usepackage{natbib}
\usepackage{hyperref}
\usepackage{comment}
\usepackage{balance}

\usepackage{xcolor}

\newcommand{\lx}{L_X}

\newcommand{\lbol}{L_{\rm bol}}

\title{Wind‑Fed Magnetic Flux and the Emergence of X‑ray Coronae in Active Galactic Nuclei}

\author[Rostom Mbarek]{
	Rostom Mbarek$^{1}$
	\\
	$^{1}$Department of Astrophysical Sciences, Princeton University, Princeton, NJ 08544, USA
}

\date{Accepted XXX. Received YYY; in original form ZZZ}
\begin{document}
	\label{firstpage}
	\pagerange{\pageref{firstpage}--\pageref{lastpage}}
	\maketitle	
	\begin{abstract}
		We propose a scenario for the onset of radiatively efficient Active Galactic Nuclei (AGN) in which circumnuclear stellar and supernova-driven winds supply mass and partially ordered magnetic flux to a compact disk reservoir around a supermassive black hole. A coherent poloidal component surviving advection to small radii strengthens inward, lowering the gas plasma beta until mean Maxwell stresses overtake MRI turbulence, compress the disk vertically, and open a magnetically dominated wedge above it. Within a quasi-force-free description of this wedge, the field self-organizes toward a critical twist at which the net radial transport changes sign, defining a magnetic saturation radius $r_c\sim10\,r_g$. The magnetic squeezing of the disk fixes the pressure balance at the interface between the disk and wedge, confining the accumulated flux at $r_c$: flux advected toward this boundary is preferentially shed, fixing $B_r\propto r^{-2}$ in the interior. From the field topology alone we then derive a normalized saturation flux $\phi \sim 50$, consistent with the MAD plateau of GRMHD simulations. The resulting compact, time-dependent saturation region is a natural candidate for the AGN X-ray corona, being collisionless, magnetized, and, at high compactness, plausibly pair-rich, with hadronic pair injection providing a conditional additional channel. Its size, variability, and spectral hardness are controlled by $r_c$, the compactness, and the magnetization. We connect these scalings to observed coronal sizes, variability, and the power partition between the corona and jet. We further propose that Little Red and Blue Dots could represent a disk-reservoir phase preceding luminous AGN activity.
	\end{abstract}

	\section{Introduction}
	
	Supermassive black holes (SMBHs) occupy a small volume fraction of their host galaxies, yet accretion-released energy and momentum can regulate gas cooling, star formation, and the thermodynamics of circumgalactic and intracluster media \citep[e.g.,][]{fabian12,king+15,hardcastle+20}. Locally, tight $M_\bullet$--bulge correlations imply that SMBH growth is coupled to galaxy assembly across cosmic time \citep[e.g.,][]{magorrian+98,ferrarese+00,gebhardt+00,kormendy+13}, while AGN feedback can offset cooling and reshape star formation on galactic and cluster scales \citep[e.g.,][]{silk+98,dimatteo+05,fabian12,king+15,harrison17}. Understanding this coupling requires linking inner-flow microphysics to the host's large-scale environment, the focus of this paper.
	
	We focus on radiatively efficient (radiative-mode) AGN, characterized by strong radiative output and prominent X-ray emission \citep{heckman+14}. These radiate at $\lambda_{\rm E}= L_{\rm bol}/L_{\rm Edd}\sim 10^{-2}$--$1$ and appear as Seyferts or quasars depending on luminosity \citep{heckman+14,padovani+17b}, powered by a radiatively efficient, geometrically thin disk \citep{shakura+73,page+74}. They are typically associated with $M_\bullet\sim 10^{6}$--$10^{8} M_\odot$ in host galaxies of stellar mass $\sim 10^{10}$ to a few $\times 10^{11} M_\odot$, with high stellar surface densities and substantial central cold-gas reservoirs and star formation \citep[e.g.,][]{heckman+14,kauffmann+03,kauffmann+09,rosario+12}. This raises the central question: what sets the onset of radiative-mode AGN?
	
	We treat this onset as coincident with the onset of X-ray emission. X-rays directly trace the compact, high-energy dissipation associated with accretion and remain comparatively resilient to host-galaxy dilution over a wide luminosity range \citep[e.g.,][]{brandt+15,padovani+17b,hickox+18,netzer15}. The hard X-ray power-law continuum is interpreted as inverse-Compton upscattering of softer seed photons by hot coronal electrons \citep{sunyaev+80,haardt+91,done+07}. Timing and spectral diagnostics localize the dominant hard-X-ray source to $\sim 10\, r_g$, where $r_g$ is the gravitational radius ($r_g= GM_\bullet/c^2$), often only a few $r_g$ above the disk via X-ray reverberation \citep[e.g.,][]{wilkins+12,kara+16}, with quasar microlensing yielding similarly compact regions of size $\sim 10\,r_g$ \citep[e.g.,][]{morgan+12}. For $M_\bullet\sim 10^6$--$10^8 M_\odot$, this corresponds to $\sim 10^{12}$--$10^{14}$ cm ($\sim 10 r_g$), and variability timescales scale with mass and accretion rate \citep[e.g.,][]{mcHardy+06,laha+25}.

    X-ray emission is also accompanied by strong optical/UV disk emission. The $\alpha_{\rm ox}$ relation linking $L_{2500\,\text{\AA}}$ and $L_{2\,{\rm keV}}$ shows that states with bright disks but weak X-rays are atypical \citep[e.g.,][]{lusso+16}. Intrinsic hard X-ray luminosity tracks other nearly isotropic power proxies. Nuclear emission in the mid infrared from reprocessed dust correlates tightly with hard X-ray luminosity corrected for absorption \citep{asmus+15}, and $[\ion{O}{iii}]\,\lambda5007$ correlates with hard X-rays over a large dynamic range \citep{heckman+05}. Thus, the onset of strong nuclear X-ray emission marks the onset of the radiative-mode.
	
	We argue in this paper that a natural physical trigger for this onset is the accumulation of dynamically important magnetic flux in the inner accretion flow. Once such fields thread the near-horizon region, they can control angular-momentum transport, energize tenuous coronal plasma, and launch winds and relativistic jets. The theoretical picture is therefore multiscale and multiphysics: the hard X-ray emission encodes particle heating, acceleration, confinement, and radiative cooling in tenuous magnetized plasma \citep[e.g.,][]{beloborodov17,groselj+24,mbarek+24,groselj+26}, while the same magnetized inner region launches disk winds and relativistic jets reaching kiloparsec scales \citep[e.g.,][]{blandford+77,blandford+82,tombes+10,veilleux+05}. Bridging these regimes spans more than twenty orders of magnitude in scale, from kinetic dissipation scales to global GRMHD structure at $r\sim$ a few--hundreds $r_g$, to the inner-kpc circumnuclear ISM. GRMHD simulations show that strong coherent magnetic flux near the hole produces magnetically dominated inner regions, powerful outflows, magnetically arrested disk (MAD) states, and efficient jet launching \citep[e.g.,][]{tchekhovskoy+11}. Local and global radiation-(GR)MHD studies further indicate that magnetic support and buoyancy shape dissipation and radiative transfer relevant for X-ray production \citep[e.g.,][]{jiang+14}. Magnetic fields thus link microphysical coronal heating and X-ray emission to macrophysical disk structure, winds, and jets.

	This raises the question of where the dynamically important flux comes from. A persistent asymmetry remains: SMBH feedback on galaxies is well studied, but how galactic environments shape the inner accretion environment and its coronal X-ray properties is less settled. We present a schematic scenario in which circumnuclear star formation supplies the magnetic flux that helps trigger radiative-mode onset. Such AGN commonly reside in galaxies with central star formation \citep{heckman+14,kauffmann+03,davies+07}, and star-forming regions and young clusters inject mass, momentum, energy, and magnetic fields via winds and supernovae \citep[e.g.,][]{chevalier+85,veilleux+05,beck15}. Gravitational torques and secular processes drive gas from kpc scales toward the nucleus, linking SMBH fueling to the same environments hosting intense star formation \citep[e.g.,][]{hopkins+10}. Consistently, wind-fed (GR)MHD simulations show that magnetized stellar winds can supply ordered fields to the inflow, building horizon-threading flux and, when sustained, driving the inner flow toward MAD \citep[e.g.,][]{ressler+20,ressler+23}.
	
	These considerations motivate our central hypothesis: winds from nuclear star-forming regions and stellar clusters supply coherent magnetic flux and magnetized plasma to the inner accretion flow, enabling a \emph{flux-saturated inner bundle} on scales of $\sim 10\,r_g$. This bundle sits inside a magnetically dominated wedge and constitutes the hard X-ray--emitting region, while its outer boundary forms a wedge saturation layer where excess twist and flux are shed.
	
	\paragraph*{Magnetic flux as the control parameter of radiative-mode onset:}

    Magnetic fields should be implicated in every high energy emission channel of AGN in the radiative-mode, providing a unifying language for AGN X-ray phenomenology. In this model, the net poloidal flux required to sustain a magnetically dominated inner saturation region, a plausible site for the corona whose microphysics is largely set by kinetic processes, is sourced and organized by nuclear regions and clusters forming stars. Strong coherent and turbulent fields are observed in galactic centers, particularly starbursts \citep[e.g.,][]{greaves+00,beck+16}, and winds driven by star formation advect magnetized plasma and flux both outward and inward, connecting star formation on kpc scales to the corona and jet launching zone at tens of $r_g$ \citep[e.g.,][]{chevalier+85,veilleux+05}. Once this large scale vertical flux reaches a disk reservoir, increasing net flux into the MAD regime increases total energy output (radiation plus outflows) \citep[e.g.,][]{avara+16,morales+18,curd+23}.

	\paragraph*{Roadmap:}
	
	Section~2 estimates the coherent magnetic flux circumnuclear stellar winds deliver to a disk reservoir and derives conditions under which the ordered component dominates angular-momentum transport. Section~3 shows how advected flux tubes could compress the disk vertically and open a magnetically dominated wedge above its surface. Section~4 analyzes wedge dynamics in a quasi-force-free framework, deriving the critical twist at which the radial electromagnetic force changes sign, the resulting magnetic saturation radius $r_c$, and the self-consistent normalized flux $\phi$. Section~5 connects these results to observed coronal sizes, spectra, variability, and corona--jet power partition, and discusses Little Blue and Little Red Dots as candidate disk-reservoir precursors. Throughout, our scaling arguments incorporate (i) observed properties of radiative-mode AGN hosts and star-forming nuclei on kpc scales, (ii) observational constraints on the inner X-ray region's size, spectrum, and variability, (iii) generic inner-flow magnetic states and outflow phenomenology from GRMHD simulations, and (iv) kinetic requirements implied by particle heating and high-energy radiative processes producing the observed hard X-ray power laws.
	
	The framework is intentionally a scaling model aimed at identifying plausible control parameters and characteristic scales, not a substitute for global or kinetic simulations. Several ingredients are treated parametrically: e.g., survival of net signed poloidal flux from $\sim 10$--$10^3\,{\rm pc}$ to $\sim 10$--$10^2\,r_g$, the relative roles of advection and turbulent diffusion in thin disks, and the 3D non-axisymmetric dynamics of interchange and flux eruption. Our goal is to articulate falsifiable correlations and a minimal topology-based organizing principle, leaving quantitative calibrations and detailed dissipation physics to dedicated simulations.

    The disk--wedge calculation is restricted to $\leq 100\,r_g$, inside the self-gravity boundary (Appendix~\ref{app:toomre}). The exterior reservoir may be clumpy or star forming \citep{goodman03,chen+23}, and the model does not require a smooth thin disk to extend from $r_c$ out to the wind-supply scale. The standard radiation-pressure branch used as a reference can be thermally unstable under the classical total-pressure stress prescription \citep{shakura+76,piran78}, and globally magnetically supported, ``flux-frozen'' solutions offer a possible alternative \citep{hopkins+24a,hopkins+24b,hopkins+24c,kaaz+25,gerling-dunsmore+26}, but neither is assumed or required here. 
	
\begin{figure}
		\centering
		\includegraphics[width=0.48\textwidth]{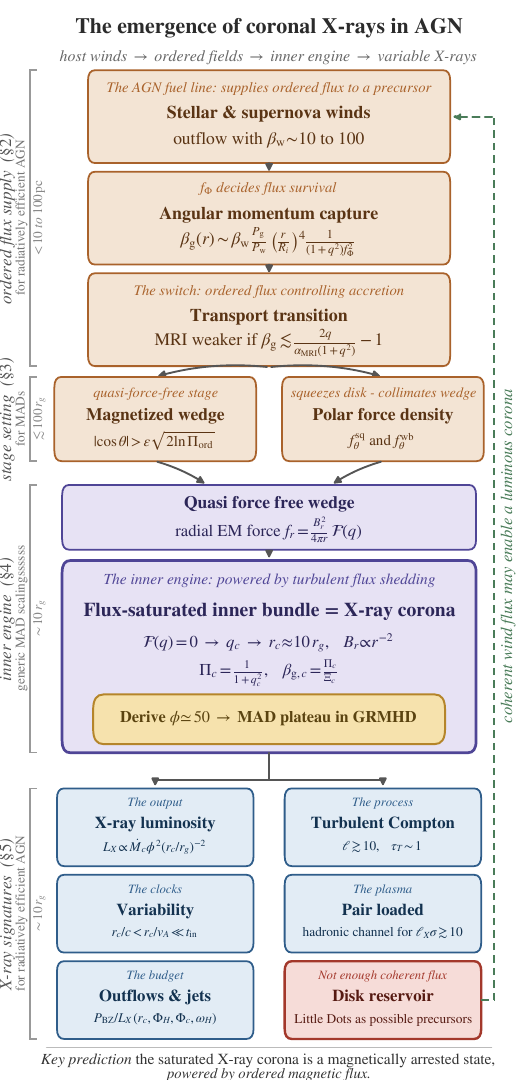}
		\caption{Schematic of the proposed pathway to radiative-mode onset. Circumnuclear star formation supplies partially ordered magnetic flux (\S\ref{sec:wind_delivery}). As the ordered component is advected inward the gas plasma beta falls and mean Maxwell stresses dominate (\S\ref{sec:angTransport}), opening a magnetically dominated wedge ($\Pi<1$, where $\Pi$ is gas-plus-radiation pressure divided by magnetic pressure) that widens as $\Pi$ falls (\S\ref{sec:ang_extent}) and squeezes the disk vertically (\S\ref{sec:psi}). Within the wedge the large-scale field self-regulates to a critical twist where the radial EM force changes sign, fixing the saturation radius $r_c\sim10\,r_g$, the interior $B_r\propto r^{-2}$, and a normalized flux $\phi\sim 50$ in the GRMHD MAD range (\S\ref{sec:dynamics}). This flux-saturated inner bundle is identified with the X-ray corona, whose topology controls the observable consequences of \S\ref{sec:observations}: an accretion-dependent $L_X$, Alfv\'enic and accretion-driven variability, hadronic pair loading, the corona--jet power partition, and X-ray-weak Little Dots as possible precursors.}
		\label{fig:roadmap}
	\end{figure}

	Fig.~\ref{fig:roadmap} summarises the proposed pathway and serves as a roadmap for the paper, running from circumnuclear flux supply (\S\ref{sec:transport}), through the magnetically dominated wedge (\S\ref{sec:wedge}), to the flux-saturated inner bundle that we identify with the X-ray corona (\S\ref{sec:dynamics}) and its observable consequences (\S\ref{sec:observations}). \\
	
	\noindent\textbf{Terminology.} We use ``magnetically dominated wedge'' for the high-latitude highly-magnetized volume in thin disk simulations; ``wedge saturation layer'' for the thin, violently regulated boundary at $r_c$ (defined below) where excess twist/flux is shed; ``flux-saturated inner bundle'' for the interior $r\lesssim r_c$ where flux is constant; and ``horizon-flux saturation'' for the MAD plateau with constant flux.

	\section{From circumnuclear star formation to inner disk magnetization}\label{sec:transport}
	
	\subsection{Wind delivery of mass and coherent flux}\label{sec:wind_delivery}
	
	\paragraph*{Magnetized winds as flux carriers}
	Star-forming regions inject energy and momentum through clustered supernovae and massive stellar winds, driving galactic outflows \citep[e.g.,][]{fielding+18}. Radio synchrotron polarimetry and Faraday rotation measurements show these outflows are magnetized with an ordered component coherent on galactic scales \citep[e.g.,][]{beck15}.
	
	The relevant launch scale $R_{\rm w}$ is the characteristic size of the star-forming region over which clustered supernovae and massive-star winds thermalize their power into a superheated outflow \citep[\S2.1 in][]{heckman+17}. In M82-like systems, the size is $R_{\rm w}\sim 10$--$10^2\,{\rm pc}$ for compact nuclear starbursts, extending to $\sim 10^2$--$10^3\,{\rm pc}$ in more extended starbursts \citep{knapen04,melioli+13,heckman+17}. Energy conservation in the \citet{chevalier+85} framework implies asymptotic wind speed $v_{{\rm hot},\infty}=(2\dot E/\dot M)^{1/2}$, giving $v_{\infty}\simeq 10^3\,{\rm km\,s^{-1}} \sqrt{\mathcal{E}/\mathcal{M}}$ (with $\mathcal{E}$ the thermalization efficiency, $\mathcal{M}$ the mass-loading) \citep{heckman+17}.
	
	The wind ram pressure at $R_{\rm w}$ is:
	\begin{equation}\label{eq:Pw}
		P_{\rm w}
		\simeq 5\times 10^{-8} 
		\, \frac{{\mathcal{M}\,\rm SFR}}{10\,M_\odot\,{\rm yr^{-1}}} \frac{v_{\infty}}{10^3\,{\rm km\,s^{-1}}} \left( \frac{R_{\rm w}}{100\,{\rm pc}} \right)^{-2}{\rm dyne\,cm^{-2}}
	\end{equation}
    For M82, the hot-wind thermal pressure $P_{\rm th}/k_{\rm B}\sim 2.5\times 10^7\,{\rm K\,cm^{-3}}$ \citep[Table~2 in][]{lacki13} lies a factor $\sim15$ below the ram pressure of Eq.~\eqref{eq:Pw}, as expected when ram pressure dominates. Multiwavelength constraints indicate strong magnetic fields $B\sim 200$--$300\,\mu{\rm G}$ in M82-like starbursts \citep{lacki13,yoast-hull+13,lopez-rodriguez+21} coherent on $\sim R_{\rm w}$ scales \citep{greaves+00}. Evaluated with this thermal pressure, the wind magnetization is $\beta_{\rm w}= 8\pi P_{\rm th}/B^2 \gtrsim 1$ for $B\sim 200$--$300\,\mu{\rm G}$ on $R_{\rm w}\sim 10$--$10^2\,{\rm pc}$ scales. The ram pressure would instead give $\beta_{\rm w}\sim10$--$30$, but the dynamically relevant plasma $\beta$ is the thermal one. In systems where the ordered field is substantially weaker or the hot phase is strongly mass-loaded, $\beta_{\rm w}$ can rise to $\sim 10^2$ \citep[e.g.,][]{beck15,heesen+11}.

	M82 itself is not a Seyfert, but serves as an illustrative pre-AGN starburst. Its unusually strong, magnetized winds should favor building an ordered nuclear flux reservoir and later Seyfert-like or quasar-like ignition. This links nuclear star formation to subsequent AGN activity without implying a guaranteed evolutionary path.
	
	\paragraph*{Angular-momentum-limited capture and flux survival}
	We model coherent magnetic flux interception by tying it to the subset of wind material that can join the nuclear inflow.  
	In the ballistic limit, a wind streamline at speed $v_{\rm rel}$ and impact parameter $b$ carries conserved specific angular momentum $b\,v_{\rm rel}$. For material to settle onto a circular orbit at radius $r$, one requires $b\,v_{\rm rel} \lesssim j_{\rm circ}(r)=\sqrt{G M_\bullet r}$ \citep{hoyle+39}, giving the maximum impact parameter for capture at $r$:
	\begin{equation}
		R_i \ \sim\ b_{\rm max}(r)\ \simeq\ \frac{\sqrt{G M_\bullet r}}{v_{\rm rel}},
	\end{equation}
	with $v_{\rm rel}\approx v_\infty$ at the base of a hot wind. For $M_\bullet=10^8M_\odot$, $v_{\rm rel}=10^3\,{\rm km\,s^{-1}}$, and $r=0.1\,{\rm pc}$, we get $R_i\simeq 0.2\,{\rm pc}$. Fields are likely coherent for $R_i \ll R_{\rm w}$ \citep{greaves+00}.

	Although Bondi--Hoyle capture need not dominate, this estimate illustrates how readily winds can supply coherent flux. In more realistic nuclear winds, upstream density gradients and non-axisymmetry inject net angular momentum into the captured stream tube \citep{edgar04,ruffert99}, ambient vorticity makes $v_{\rm rel}$ stochastic \citep{krumholz+05,krumholz+06}, and MHD effects can alter both mass throughput and net coherent poloidal flux that survives advection \citep{igumenshchev+02}. A coherence survival factor $f_\Phi$ bundles geometric suppression and field cancellation. This factor need not vanish during inward advection, as shown by the wind-fed Sgr~A* simulations of \citet{ressler+20}, a proof of concept for coherent flux survival, albeit in a different accretion regime.

	\subsection{Angular momentum transport by ordered fields}\label{sec:angTransport}
	
	A net poloidal component can accumulate at small radii when inward advection overcomes outward transport, including turbulent diffusion \citep{spruit+05,guilet+12,guilet+13}. Global GRMHD simulations show that, once the inner flow is strongly magnetized, angular-momentum transport is largely set by the net ordered poloidal flux delivered to the hole. As the normalized horizon-threading flux $\phi$ approaches saturation, excess magnetic flux accumulates outside the horizon, partially arresting the inflow and establishing a MAD state \citep{tchekhovskoy+11,narayan+12}. In MAD regions, the standard MRI\footnote{MRI is a major source of disk turbulence and stress in ionized, weakly magnetized, and differentially rotating plasmas \citep{balbus+91,balbus+98}} can be suppressed, while accretion proceeds through non-axisymmetric motions and magnetic stresses whose mean and fluctuating contributions depend on the flow structure \citep{mckinney+12}. Recent global simulations confirm these behaviors \citep{liska+19,liska+20}, and time-dependent MAD simulations exhibit episodic flux loss and flare-like events associated with plasmoid-mediated flux ejection \citep{ripperda+22}.


	For a net ordered field threading the inner disk, we write $\boldsymbol{B}=\langle\boldsymbol{B}\rangle+\delta\boldsymbol{B}$ and focus on Maxwell stresses controlled by $\langle B_r\rangle$ and $\langle B_\phi\rangle$. Azimuthal averaging separates the radial Maxwell stress into mean-field and fluctuating contributions:
	\begin{equation}
		\langle T_{r\phi}\rangle
		= \frac{\langle B_r\rangle\langle B_\phi\rangle+\langle \delta B_r \delta B_\phi\rangle}{4\pi}
		= T_{r\phi}^{\rm ord}+T_{r\phi}^{\rm tur}
		\label{eq:maxwell_stress_split}
	\end{equation}
	with $\alpha_{\rm tur}= T_{r\phi}^{\rm tur}/(P_{\rm g}+P_{\rm B})|_{\rm mid}$ representing the fluctuating Maxwell stress, associated primarily with MRI turbulence \citep[e.g.,][]{davis+10,bai+13}.
	
	We define the twist $q(r)= |\langle B_\phi\rangle/\langle B_r\rangle|$ and the ordered-field stress parameter $\alpha_{\rm ord}= T_{r\phi}^{\rm ord}/(P_{\rm g}+P_{\rm B})$, whose magnitude is $|\alpha_{\rm ord}| \simeq 2q/[(1+q^2)(1+\beta_{\rm g})]$, with $\beta_{\rm g}=P_{\rm g}/P_{\rm B}$ the gas plasma beta. Radiation pressure $P_{\rm rad}$ supports the disk vertically but does not resist shear, so it is excluded from the transport stresses. It is carried by the factor $\Xi=1+P_{\rm rad}/P_{\rm g}\ge1$ (Appendix~\ref{app:prad_pgas}) and enters only in the vertical-support physics of \S\ref{sec:wedge}. The condition $|\alpha_{\rm ord}|\gtrsim \alpha_{\rm tur}$ gives:
		\begin{equation}
		\frac{2q}{(1+q^2)(1+\beta_{\rm g})}\gtrsim \alpha_{\rm tur}
		\Rightarrow
		\beta_{\rm g}\lesssim 
		\frac{2q}{\alpha_{\rm tur}(1+q^2)}-1
		\label{eq:betaLS_threshold_alphaMRI}
	\end{equation}
	The right-hand side is maximized at $q=1$, giving $\beta_{\rm g}\lesssim (\alpha_{\rm tur})^{-1}-1$. For $\alpha_{\rm tur}\sim 10^{-2}$--$10^{-1}$ and $q\sim\mathcal{O}(1)$, this corresponds to $\beta_{\rm g}(r)\lesssim\mathcal{O}(10\text{--}100)$. The transport transition is therefore controlled by the gas plasma beta.

    Dominance of the ordered Maxwell stress over the MRI stress (Eq.~\eqref{eq:betaLS_threshold_alphaMRI}) governs angular-momentum transport, but does not by itself guarantee that inward flux advection beats turbulent diffusion. Vertically averaged isotropic-transport models favor outward diffusion \citep{lubow+94}, whereas vertical structure and fast surface-layer inflow can enhance inward transport \citep{zhu+18}. 
	
	In ideal MHD, the poloidal flux $\Phi_p$ threading a bundle of streamlines is preserved as it is advected inward. While the field delivered at large radii is predominantly vertical, accretion progressively bends it into a more radial configuration, so we treat $B_p $ and $ B_r$ interchangeably. In this limit, weak lateral mixing conserves the bundle's solid-angle footprint $\Delta\Omega$, so $\Phi_p\simeq B_p\, r^2\Delta\Omega$ gives $B_p\propto r^{-2}$ (Appendix~\ref{app:flux-scaling}). Departures from these idealizations can modify the precise radial slope of $B_p$, but this qualitative picture is likely to stay relevant. Mass conservation gives $\rho\propto r^{-2}$, and for an adiabatic gas $P_{\rm g}\propto\rho^{\Gamma_{\rm ad}}\propto r^{-2\Gamma_{\rm ad}}$, whereas $P_{\rm B}\propto r^{-4}$. Hence $\beta_g\propto r^{4-2\Gamma_{\rm ad}}$. For $\Gamma_{\rm ad}<2$, $P_{\rm B}$ grows more rapidly than $P_{\rm g}$ under transverse compression, so $\beta_g$ decreases inward and the ordered component becomes progressively more magnetically dominated.
	Because the transport handover of Eq.~\eqref{eq:betaLS_threshold_alphaMRI} is set by $\beta_{\rm g}$, this adiabatic scaling governs it directly, with no reference to radiation.
	
	The coherent flux delivered to the disk reservoir is set by the poloidal captured raw flux suppressed by the survival factor $f_\Phi$, $\Phi_{\rm ord}(r)\simeq f_\Phi\,\pi R_i^2 B_{{\rm w},p}$, implying $B_{{\rm ord},p}(r)\sim f_\Phi B_{{\rm w},p}(R_i/r)^2$. With differential rotation winding giving $B_{\rm ord}^2(r)\simeq (1+q^2)B_{{\rm ord},p}^2(r)$, the advected gas plasma $\beta$ at radius $r$ is:
	\begin{equation}
		\beta_{\rm g}(r)\sim
		\beta_{\rm w} \frac{P_{\rm g}(r)}{P_{\rm th}}
		\left(\frac{r}{R_i}\right)^4
		\frac{1}{(1+q^2)f_\Phi^2}
		\label{eq:betaLS_from_betaw}
	\end{equation}

	The corresponding support ratio $\Xi\beta_{\rm g}$ carries the impact of radiation, where vertical support matters ($\Xi\to1$ in the gas limit, $\Xi\gg1$ under radiation domination). The model assumes externally supplied net poloidal flux wound toroidally by differential rotation, as in net-flux disk simulations \citep{salvesen+16,zhu+18}. It does not require a self-sustaining toroidal field, though such states offer an alternative route to magnetic support \citep{gaburov+12,squire+25}. A large $\beta_{\rm g}$ leaves transport MRI-dominated and delays advection, but as net flux accumulates inward $\beta_{\rm g}$ drops, the mean Maxwell stress strengthens, and the ordered field increasingly controls the torque, driving the inner flow toward MAD \citep[e.g.,][]{ressler+23}. This outcome is not expected if $\beta_{\rm g}$ remains too large.
	
	\section{Formation of the magnetically dominated wedge}\label{sec:wedge}
	
	\subsection{Angular extent of the magnetically dominated wedge}\label{sec:ang_extent}
	
	Let $\epsilon= H/r$ denote the disk aspect ratio, where $H$ is the scale height such that $\epsilon\ll 1$ for thin disks. For a vertically isothermal disk the gas pressure follows $P_{\rm g}(z)\propto\exp[-z^2/(2H^2)]$ \citep{shakura+73,pringle81}. With $z=r\cos\theta$,
  \begin{equation}
      P_{\rm g}(r,\theta)\simeq P_{\rm g}(r)
      \exp\!\left[-\frac{\cos^2\theta}{2\epsilon^2}\right].
  \end{equation}
    The vertical shape is thus set by the gas. The support pressure is $P_{\rm sup}=P_{\rm g}+P_{\rm rad}=\Xi P_{\rm g}$ ($\Xi$ as in \S\ref{sec:angTransport}), taken slowly varying across the thin wedge base
  (Appendix~\ref{app:prad_pgas}), so $P_{\rm sup}$ inherits the same Gaussian to leading order in the gas-supported base. For $\Xi\gg1$ the radiation-supported profile is shallower and this is a reference approximation (see below). 
  
  We define the support-to-magnetic ratio $\Pi_{\rm ord}= P_{\rm sup}/P_{\rm B}=\Xi\beta_{\rm g}$, and if the coherent mean field varies weakly with $\theta$, then $B^2(r,\theta)\simeq B_{\rm ord}^2(r)$ and $\Pi(r,\theta)\simeq
  \Pi_{\rm ord}(r)\exp[-\cos^2\theta/(2\epsilon^2)]$. Magnetic domination
  ($\Pi(r,\theta)<1$) requires:
	\begin{equation}\label{eq:ff_wedge_condition}
		|\cos\theta|>\epsilon\sqrt{2\ln\Pi_{\rm ord}(r)}.
	\end{equation}
	where the wedge opening is radius-dependent. This Gaussian form holds for a gas-supported base. When radiation dominates ($\Xi\gg1$), $P_{\rm sup}$ follows a polytrope, $P_{\rm sup}\propto\rho^{1+1/n_{\rm pol}}$, and magnetic domination instead requires (Appendix~\ref{app:rad_wedge})
\begin{equation}
    |\cos\theta|>\frac{z_0}{r}\left(1-\Pi_{\rm ord}^{-1/(n_{\rm pol}+1)}\right)^{1/2},
\end{equation}
with $z_0$ the zero-pressure surface and $n_{\rm pol}=3$ for a radiation-dominated layer. We retain the Gaussian form as the gas-supported reference used throughout.
    
    Physically, the wedge opening is a manifestation of the transport transition described in \S\ref{sec:angTransport}. Because $\Pi_{\rm ord}$ decreases inward, while $P_{\rm g}$ falls as a Gaussian with height and the ordered magnetic pressure $B^2$ is nearly independent of $\theta$, the $\Pi=1$ surface moves closer to the disk midplane at smaller radii. This exposes an increasingly broad magnetically dominated wedge above the disk.
	
\begin{figure}
		\centering
		\includegraphics[width=0.45\textwidth,clip=false,trim= 0 0 0 0]{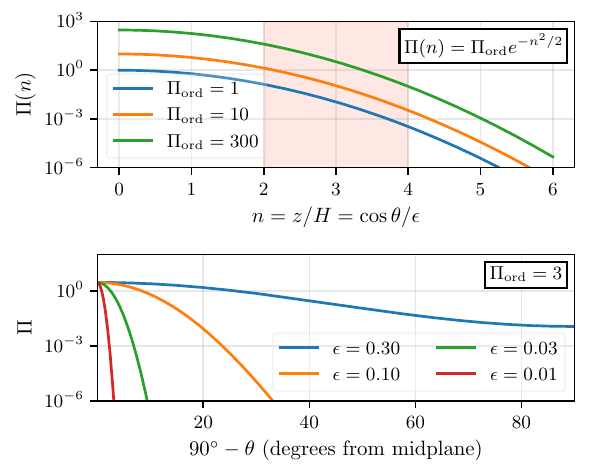}
		\caption{Support-ratio ($\Pi$) structure in the disk atmosphere for
        $\Pi(n_H)=\Pi_{\rm ord}e^{-n_H^2/2}$ with $n_H= z/H=\cos\theta/\epsilon$ for $\Xi \simeq 1$. Top: $\Pi$ versus scale height for several $\Pi_{\rm ord}$ (transition to magnetic dominance is confined to $n_H\sim2$--$3$). Bottom: the same profile mapped to polar angle for fixed $\Pi_{\rm ord}$ and different disk aspect ratios $\epsilon$. The same range in $n_H$ corresponds to a narrow angular wedge in thick disks and a broad one in thin disks. }
		\label{fig:beta}
	\end{figure}

	Overall, larger $\beta_{\rm w}$ raises $\Pi_{\rm ord}$ and pushes the $\beta<1$ surface to larger $|\cos\theta|$. More efficient flux transport (larger $f_\Phi$) or stronger winding (larger $q$) lower $\Pi_{\rm ord}$ and open the wedge toward the disk surface. A larger wedge volume occurs for smaller values of  $\cos\theta$, and enhanced mass loading shrinks or eliminates such a wedge. It is useful to recast this condition in units of the local scale height by defining $n_H= z/H=\cos\theta/\epsilon$, so that
	\begin{equation}
		\Pi(n_H) \simeq \Pi_{\rm ord}(r)\,\exp\!\left[-n_H^2/2\right],
		\label{eq:beta_n}
	\end{equation}
	In this sense, and as shown in Figure~\ref{fig:beta}, geometric differences between thick and thin disks are absorbed into the mapping $\cos\theta=\epsilon \, n_H$. For reasonable $1 \leq \Pi_{\rm ord} \leq 100$, the magnetically important transition where support and magnetic pressures become comparable is confined to a narrow range in height, typically $n_H\sim2$--$3$. This follows from the approximately Gaussian vertical stratification, $P_{\rm gas}\propto e^{-n_H^2/2}$, which drives $\Pi$ down if the ordered magnetic pressure varies more slowly with height. In angular terms, the same range corresponds to a broad polar cone in thin disks ($\epsilon\ll0.1$) but to a narrow one in thicker disks ($\epsilon\sim0.3$). For very thin disks, the isothermal Gaussian atmosphere is less reliable for $n_H\gtrsim 1$, and in a radiation-supported base ($\Xi\gg1$) the vertical profile departs from Gaussian (Appendix~\ref{app:rad_wedge}), so the $n_H\sim2$--$3$ transition is a gas-supported approximate reference. 
	
	In summary, wedge formation requires: (i) a coherent mean field surviving transport without cancellation; (ii) differential rotation winding the poloidal field into a toroidal component, also amplifying the ordered field by $(1+q^2)^{1/2}$; (iii) transverse compression of advected flux tubes lowering $\Pi_{\rm ord}$; (iv) vertical stratification based on $\epsilon$; and (v) weak high-latitude mass loading. Since wedge formation is governed by $\epsilon$, $f_\Phi$, $q$, and the transverse pressure profile rather than $\beta_{\rm w}$ directly, the initial plasma $\beta$ of the injected poloidal field in GRMHD experiments has limited predictive power for whether the flow reaches different MAD states. What matters is whether and how a net signed poloidal flux survives advection to small radii. 
	
	The polar cone defined by $0\le\theta<\theta_d(r)$ is bounded by the $\beta=1$ surface, with opening $\theta_d=\arccos\!\big[\epsilon\sqrt{2\ln\Pi_{\rm ord}(r)}\big] \leq \pi/2$ from Eq.~\eqref{eq:ff_wedge_condition}. Note that $\epsilon\sqrt{2\ln\Pi_{\rm ord}}\to0$ or $\theta_d \to \pi/2$ occurs for a vanishingly thin disk. Finite disk thickness and residual support pressure ($\Pi_{\rm ord}>1$) always push $\theta_d$ below $\pi/2$. Near the black hole, the horizon-threading polar funnel of angle $\theta_f(r)$ excludes the innermost polar region, so the magnetically dominated wedge is restricted to $\theta_f(r)<\theta<\theta_d(r)\le\pi/2$.

\subsection{Disk squeezing from advected ordered fields}\label{sec:psi}

After establishing the angular structure of the wedge, we now ask how the same inward-advected flux can modify the disk vertical structure beneath it. The external confining stress is set by the magnetic-pressure contrast across the disk--wedge interface, $\Delta P_B\simeq B_r^2/(8\pi)=P_B/(1+q^2)$ (see \S\ref{sec:polar}). Including gas and radiation in $P=P_{\rm g}+P_{\rm rad}$, and replacing the standard boundary condition $P(H)\simeq 0$ with $P(H)\simeq\Delta P_B$, the approximate integrated hydrostatic balance gives
\begin{equation}
    P_{\rm mid}-\Delta P_B\simeq\Sigma\Omega_K^2H/2
    \label{eq:pmid}
\end{equation}
where $\Sigma$ is the disk surface density and $\Omega_K$ the Keplerian frequency. Let $P_{\rm mid,0}=P_{{\rm g},0}+P_{{\rm rad},0}\simeq\Sigma\Omega_K^2H_0/2$ describe the disk without magnetic confinement. Assuming $P_{\rm mid}H\simeq P_{\rm mid,0}H_0$ during the vertical adjustment yields
\begin{equation}
    H\simeq H_0\left(\sqrt{1+\Psi^2}-\Psi\right),
    \qquad
    \Psi=\frac{\Delta P_{\rm B}}{2P_{\rm mid,0}},
    \label{eq:H_solution}
\end{equation}
recovering $H\to H_0$ as $\Psi\to0$ and $H\simeq H_0/(2\Psi)\ll H_0$ for $\Psi\gg1$. This pressure-response approximation includes radiation through $P_{\rm mid,0}$, and in the isothermal gas-supported limit, $H_0=c_s/\Omega_K$ and $\Psi=\Delta P_{\rm B}/(\Sigma\Omega_Kc_s)$, with $c_s$ the gas sound speed. Defining $\epsilon_0=H_0/r$, the magnetically squeezed aspect ratio is
\begin{equation}
    \epsilon(r)=\frac{H}{r}
    \simeq\epsilon_0(r)\left(\sqrt{1+\Psi^2}-\Psi\right).
    \label{eq:epsilon}
\end{equation}
Since flux freezing gives $P_{\rm B}\propto r^{-4}$, one has $\Psi\propto r^{w+u-5/2}$ for $\Sigma\propto r^{-w}$, $c_s\propto r^{-u}$ and $\Omega_K \propto r^{-3/2}$. Because magnetic domination above the midplane steepens inward, it is natural to require $\Psi$ to increase toward small radii, which implies $w+u<5/2$. At the same time, steady accretion at fixed
	$\dot M_{\rm tot} = 3\pi\nu\Sigma$, with $\nu\sim\alpha c_s H$, requires
	\begin{equation}
		\nu\Sigma\propto r^{3/2-2u-w}
		\left(\sqrt{1+\Psi^2}-\Psi\right)
	\end{equation}
	to be approximately constant with radius. Since the second factor approaches unity at large radii, steady thin-disk accretion enforces $w+2u\simeq3/2$ asymptotically. Magnetic squeezing modifies the local exponent through the $\Psi$-dependent factor, but does not affect this large-radius balance. For reference, the gas-pressure, electron-scattering Shakura--Sunyaev scalings \citep{shakura+73} give $w=3/5$ and $u=9/20$, so $w+2u=3/2$ and $\Psi\propto r^{-1.45}$. The Kramers-opacity scaling gives $w=3/4$ and $u=3/8$, again with $w+2u=3/2$, and $\Psi\propto r^{-1.375}$. In both cases $w+u<5/2$, so $\Psi$ increases inward and the advected field squeezes the disk progressively more strongly toward small radii.


    For the standard radiatively efficient, UV-emitting reference disk, Appendix~\ref{app:thin_disk} gives $H/r\simeq0.29\,\lambda_{\rm E}$ and $\Sigma\sim10^3$--$10^4\,{\rm g\,cm^{-2}}$ at $r=10\,r_g$, dense enough that photons escape well before inflow. The disk may instead be truncated near $r_c$, with magnetic squeezing potentially contributing to this transition. However, this model does not predict the truncation radius, since it treats the quasi-force-free wedge and invokes the disk locally, for pressure balance at the interface.

	\paragraph*{When does squeezing become dynamically relevant?}
    
    For the uncompressed atmosphere, $P_{\rm sup}(z_{\rm int})=P_B$ defines the height $z_{\rm int}=n_{\rm int}H$, where $n_{\rm int}=\sqrt{2\ln\Pi_{\rm ord}}$ (Eq.~\eqref{eq:ff_wedge_condition}). The fractional gas column above this height provides a diagnostic for the onset of squeezing (Appendix~\ref{app:column_fraction}):
    \begin{equation}
    \frac{\Sigma(>z_{\rm int})}{\Sigma_{\rm total}}
    =\frac{1}{2}\operatorname{erfc}
      \!\left(\sqrt{\ln\Pi_{\rm ord}}\right),
    \label{eq:column_fraction}
    \end{equation}
	which vanishes as $\Pi_{\rm ord}\to\infty$. When
	$\Pi_{\rm ord}\gg 1$, the interface resides high in the
	atmosphere and only a thin skin of gas sits in the squeezed layer. The disk interior is likely not qualitatively restructured by the ordered field. For the squeezing layer to extend into a dynamically significant portion of the column, the interface must descend further, requiring
	\begin{equation}
		n_{\rm int}=\sqrt{2\ln\Pi_{\rm ord}(r)}\lesssim\mathrm{few}
		\quad\Longleftrightarrow\quad
		\Pi_{\rm ord}(r)\lesssim\mathcal{O}(\mathrm{few}).
		\label{eq:squeezing_onset}
	\end{equation}
	Under the same reference scalings, $\Pi_{\rm ord}\propto r^{5/2-w-u}/(1+q^2)$, so the interface position depends on both stratification and winding. For $\Pi_{\rm ord}=2$, the interface is at $n_{\rm int}\simeq1.2$ and about $12\%$ of the total column lies above it on one side. At $\Pi_{\rm ord}=1$, the interface reaches the midplane and that fraction is $50\%$. The squeezing layer therefore extends into a substantial part of the disk body as $\Pi_{\rm ord}$ approaches unity.
    For the uncompressed reference atmosphere, $\Psi\simeq[2(1+q^2)\Pi_{\rm ord}]^{-1}$. The column fraction above equipartition measures the depth of magnetic domination, while substantial compression requires $\Psi\sim1$ and depends on the radial-field fraction.

	\paragraph*{Linear response of the whole column.}
	The threshold in Eq.~\eqref{eq:squeezing_onset} identifies the onset of a squeezing regime, where the wedge interface descends into the disk body and the ordered field occupies a dynamically significant portion of the column. At larger radii, $\Pi_{\rm ord}\gg 1$ places the interface high in the atmosphere, but the disk still feels $\Delta P_{\rm B}$ as an external boundary pressure through the integrated momentum balance of Eq.~\eqref{eq:pmid}: for $\Psi\ll 1$, Eq.~\eqref{eq:epsilon} gives $H/H_0\simeq 1-\Psi\simeq 1-\frac{1}{2(1+q^2)\Pi_{\rm ord}}$, so the whole column contracts by a fraction $\sim\Psi$ and the thermal emission brightens correspondingly. The distinction is therefore between mild column-compression (weak squeezing wherever $\Psi$ is non-negligible and capable of enhancing UV disk emission) and structural restructuring of the disk interior (strong squeezing, requiring $\Psi\gtrsim1$. In the strong regime, additional heating from the compression may be radiated efficiently, allowing the disk to relax to an even smaller $H$.
	Radiation-MHD simulations demonstrate how radiation and magnetic fields jointly shape disk structure and turbulence \citep[e.g.,][]{jiang+14,secunda+24}.
	
	\section{Dynamics inside the magnetically dominated wedge}\label{sec:dynamics}
	
	\subsection{Flux regulation and the magnetic saturation radius}\label{sec:topology}
	
	Motivated by the formation of regions with $\beta\ll 1$ and magnetization $\sigma= B^2/(4\pi \rho c^2)\gtrsim 1$ in MAD GRMHD simulations \citep[e.g.,][]{mckinney06,ball+18,chael+19,ressler+21}, we model the global dynamics of the magnetically dominated wedge in a quasi-force-free (FF) framework, appropriate when magnetic stresses dominate both inertia and thermal pressure. In this limit, turbulent fluctuations are subdominant and $q= B_\phi/B_r$ refers to the magnitudes of the large-scale coherent field ratio. We use the quasi-FF approximation to capture the large-scale structure of EM stresses, excluding microphysical details and regions where $\sigma \ll 1$.

    For $B_\theta=0$, radial flux transport depends on both angular momentum exchange, governed by $T_{r\phi}=B_rB_\phi/(4\pi)$, and magnetic pressure and tension, described by $T_{rr}=(B_r^2-B_\phi^2)/(8\pi)$, $T_{\theta\theta}=-(B_r^2+B_\phi^2)/(8\pi)$, and $T_{\phi\phi}=(B_\phi^2-B_r^2)/(8\pi)$. Under axisymmetry, the radial magnetic force is the Maxwell divergence $f_r^{B}=(\nabla\cdot\mathbf{T})_r=r^{-2}\partial_r(r^2T_{rr})-(T_{\theta\theta}+T_{\phi\phi})/r$, combining pressure, tension, and curvature. To represent the additional influence of accretion, we introduce $r^{-2}\partial_r(r^2T_{r\phi})$ as an effective transport contribution. For a shell of fixed solid angle, the radial faces have area $A=r^2\Delta\Omega$, so this term measures the difference in shear traction multiplied by face area, per unit shell volume. It captures how the transmitted shear stress varies between neighbouring shells while accounting for their increasing area. Radial gradients of the magnetic shear stress can drive radial motion. This follows from mass and angular momentum conservation, which relate the inflow speed to derivatives of the magnetic stresses, including the $r\phi$ term \citep{rothstein+08}. This inflow has also been used to compute inward flux advection in thin disks threaded by a coherent magnetic field \citep{cao+13,begelman24}. We combine this shear gradient term with $f_r^{B}$ to define $f_r$, which represents radial transport in this model:
\begin{equation}
    f_r =
    \frac{1}{r^2}\partial_r\!\left[r^2(T_{rr}+T_{r\phi})\right]
    -\frac{T_{\theta\theta}+T_{\phi\phi}}{r}.
\end{equation}
The zero of $f_r$ defines the threshold for transport balance. Parameterizing the coherent large-scale field as power laws with $B_r\simeq B_0(r/r_0)^{-p}$ and $B_\phi\simeq B_0(r/r_0)^{-g}$ yields $f_r=\frac{B_r^2}{4\pi r}\mathcal{F}(q)$ such that:
\begin{align} 
		\mathcal{F}(q)=\left(2-p\right)+(2-p-g) q+\left(g-1\right)q^2 \label{eq:Fq}
\end{align}
equivalently
\begin{align} 
f_r(r)
&= \frac{B_0^2}{8\pi r_0}
\left(\frac{r}{r_0}\right)^{-(p+g+1)}
\Bigg[
(4-2p)\left(\frac{r}{r_0}\right)^{-(p-g)}
\nonumber\\
&\qquad
+(2g-2)\left(\frac{r}{r_0}\right)^{-(g-p)}
+(4-2p-2g)
\Bigg].
\label{eq:fr_full}
\end{align}
We define a critical radius $r_c$ by $f_r(r_c)=0$, marking the transition between net inward and net outward EM forcing. We restrict to $p,g\geq 1$, ensuring both $B_r$ and $B_\phi$ decline at least as $1/r$.
The shared amplitude $B_0$ sets $q_0=1$ at a footpoint $r_0$ where $B_\phi\simeq B_r$, plausibly near or below the outer light cylinder (a few $r_g$ for high spin), where winding saturates at $q\sim c/v_A\sim1$. This footpoint does not coincide with the horizon (see Appendix~\ref{app:qH}).

The slopes $p$ and $g$ are not independent since the toroidal component is produced by winding the poloidal field: $g=p-d\ln q/d\ln r$. Balancing shear generation of $B_\phi$ against removal on a timescale $t_{\rm loss}$ gives $q\sim \Omega t_{\rm loss}$ in steady state. For Alfv\'enic removal, $t_{\rm loss}\sim r/v_A$, so $q\sim \Omega r/v_A$. If $\Omega\simeq\Omega_K$ and $v_A$ is roughly constant, then $q\propto r^{-1/2}$ and $g$ exceeds $p$ by $\simeq 1/2$. If instead $v_A$ increases inward, as expected for a strengthening ordered field, Alfv\'enic removal counteracts shear-driven winding and drives $q$ downward with decreasing radius, pushing toward $g\simeq p$ or even $g<p$.

	\subsection{Twist-regulated forcing and stress saturation}
	
\begin{figure}
		\centering
		\includegraphics[width=0.45\textwidth,clip=false,trim= 0 0 0 0]{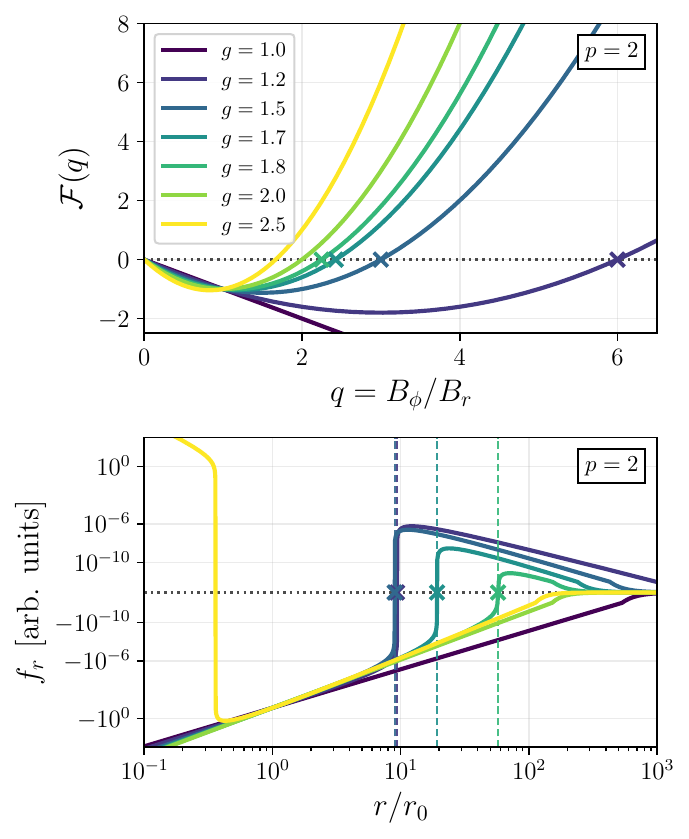}
		\caption{Upper panel: Solutions to $\mathcal{F}(q)=0$ for different slopes $g$. Lower panel: Solutions to $f_r=0$ normalized to $r_0$, with dashed lines marking the saturation region $r_c/r_g \simeq (r_0/r_g)\,q_c^{1/(p-g)}$, where $r_0 \gtrsim r_g$. In both panels, crosses mark the positive root $q_c$ (and corresponding $r_c$) and are shown for $g<p$, i.e. when a saturation radius $r_c/r_0>1$ exists.}
		\label{fig:combined}
	\end{figure}

	In the upper panel of Figure~\ref{fig:combined}, we plot solutions of Eq.~\eqref{eq:Fq} for fixed $p=2$ (see \S\ref{sec:flux_sat}) and several threading-dependent slopes $g$. The function $\mathcal{F}(q)$ has a single minimum at $q_\star=\frac{p+g-2}{2(g-1)}\sim\mathcal{O}(1)$. For $q<q_\star$, $\mathcal{F}'(q)<0$, so increasing twist drives $\mathcal{F}(q)$ (and $f_r$) more negative. The inward force strengthens up to $q_\star$, where $\mathcal{F}$ reaches its minimum and the force has its greatest inward magnitude. At $q=q_c$, $f_r$ vanishes, and for $q>q_c$ it becomes outward. The system therefore cannot sustain arbitrarily large twist. Once the inward branch saturates near $q_c$, further winding is relieved by transport and $r_c$ marks a saturation radius.
	
	The same twist $q$ controls angular momentum transport. The $T_{r\phi}$-stress is $\frac{|B_rB_\phi|}{4\pi}=\frac{B_r^2}{4\pi} q$, so increasing $q$ at fixed $B_r$ increases the magnitude of the magnetic shear stress. For $q<q_c$ the force is inward (confining), whereas for $q>q_c$ the field tends to inflate and shed excess toroidal flux. The twist self-adjusts toward $q\simeq q_c$, and the Maxwell stress becomes self-limited to $|T_{r\phi}(r_c)|\sim \frac{B_r^2(r_c)}{4\pi} q_c(p,g)$ at the saturation radius.

	\subsection{Existence and intermittency of saturation for different slopes}\label{sec:rc_def}
	
    For $g>1$ the coefficient of $q^2$ in $\mathcal{F}(q)$ is positive, so $\mathcal{F}(q)$ is convex and admits a single positive root:
	\begin{equation}
		q_c
		=\frac{(p+g-2)+\sqrt{(p+g-2)^2-4(g-1)(2-p)}}{2(g-1)}
		\label{eq:qc_general}
	\end{equation}
	For $p=2$ one has $\mathcal{F}(1)=3-2p=-1<0$ while $\mathcal{F}(q)\to +\infty$ as $q\to\infty$, so $q_c>1$. We therefore restrict to $1<g<p$, where a compact saturation radius exists. The limiting case $g\to1$ removes the quadratic term and admits no finite $r_c$.
	

The power-law fields imply $q(r)=q_0(r/r_0)^{p-g}$, normalized at the footpoint $r_0$ where $B_\phi\simeq B_r$, so $q_0\simeq1$. Imposing $q(r_c)=q_c$ then gives:
	\begin{equation}\label{eq:qc_rc}
		q_c=q_0\left(\frac{r_c}{r_0}\right)^{p-g}
		\quad\Longrightarrow\quad
		\frac{r_c}{r_g}\simeq \frac{r_0}{r_g}\,q_c^{1/(p-g)},
	\end{equation}
with $r_0 \gtrsim r_g$. For the fiducial normalizations below we take $r_0\simeq r_g$, so $r_c/r_g\simeq q_c^{1/(p-g)}$ and the $(r_0/r_g)$ prefactor is absorbed into the order-of-magnitude estimates. A footpoint at $r_0=\mathcal{O}(\mathrm{few})\,r_g$ would raise $r_c$ and $\phi\propto(r_c/r_g)^{3/4}$ accordingly.
For $p>g$, $q$ grows outward from the adopted value
$q_0\simeq1$ at $r_0$ (Appendix~\ref{app:qH}) and reaches $q_c$ at $r_c>r_0$.
    
For $g>p$, the exterior power law makes $q$ decrease outward
from the adopted value $q_0\simeq1$ at $r_0$
(Appendix~\ref{app:qH}). If $q_0<q_c$, no saturation radius
occurs at $r\ge r_0$ within this profile
(Appendix~\ref{app:ivt}). If another process raises $q$ to exceed $q_c$ somewhere outside the horizon, the expected outcome is saturation. Since $\mathcal{F}(q)$ is convex with a single positive root, any monotonic inward increase of the twist crossing $q_c$ places smaller radii on the expelling ($\mathcal{F}(q)>0$) branch.
	
In the magnetically dominated wedge, black-hole spin could indirectly contribute to the twist $q$ through ergospheric coupling rather than direct horizon-threading. This could be related to the ergomagnetospheric picture of \citet{blandford+22}, in which an ergospheric current system transmits spin energy outward.

	\subsection{Flux saturation for $r\leq r_c$}\label{sec:flux_sat}
	
	For $p>g$, at $r=r_c^{+}$, $\mathcal{F}>0$ and
	$f_r=K(r)\Phi^2\mathcal{F}(q)$ with $K(r)=\frac{1}{4\pi^3}r^{-5}>0$
	and $\Phi(<r)=\pi r^2 B_r$. For a small, rapidly-applied flux
	perturbation at fixed $q$,
	\begin{align}
		\Big(\frac{\partial f_r}{\partial \Phi}\Big)_{q}\bigg|_{r_c^{+}}
		= 2K\,\Phi\,\mathcal{F}\bigg|_{r_c^+} > 0,
	\end{align}
	since $K>0$, $\Phi>0$, and $\mathcal{F}>0$ at $r_c^{+}$. Any
	additional flux $\delta\Phi$ advected in from larger radii therefore
	produces
	\begin{align}
		\delta f_r(r_c^{+})
		\;\approx\;
		\Big(\frac{\partial f_r}{\partial \Phi}\Big)_{q}\delta\Phi
		=2K\,\Phi\,\mathcal{F}\,\delta\Phi \;>\; 0,
	\end{align}
	so extra flux arriving just outside $r_c$ increases the outward $f_r$ precisely where the profile is already peaked at $r_c^{+}$ (see Figure~\ref{fig:combined}), and the response is immediate outward transport that expels the surplus. The only steady configuration has $\Phi(<r)=\Phi(r_c)$ for $r<r_c$, making $r_c$ a \textit{magnetic saturation radius} and the interior a
	\textit{flux-saturated inner bundle} with $B_r\propto r^{-2}$ (i.e.,
	$p=2$). Figure~\ref{fig:combined} shows that this saturated region sits at $r_c\sim 10\,r_g$ over a range of $g$.
	
	The flux-saturated inner bundle should not be conflated with the horizon-threading flux anchoring the relativistic jet. In MAD simulations, the polar funnel is organized into open, BH-threaded field lines with $\sigma\gg 1$, whereas the disk and wedge sit at much lower magnetization. 
    The transition from jet ($\sigma\gg 1$) to disk ($\sigma\ll 1$) is mediated by a $\sigma\sim\mathcal{O}(1)$ buffer zone in the funnel-wall and wedge layers, the focus of this model.

    The magnetic saturation radius $r_c$ need not coincide with thermal disk truncation. If the optically thick disk extends through $r_c$, its photosphere anchors the wedge base, while only a small fraction of the inflow loads the magnetized phase. Truncation near $r_c$ exposes a larger solid angle but need not brighten the corona. The net effect on the seed-photon supply and inverse-Compton cooling is uncertain. Cold clumps may coexist with the hot, magnetically dominated corona, as in the three-dimensional simulations of \citet{liska+22}. The current model does not resolve this multiphase structure, nor does it predict the truncation radius. A disk that stays optically thick through $r_c$ and one truncated near $r_c$ are both allowed realizations, differing in geometry and UV seed supply.

	\subsection{Polar force structure}\label{sec:polar}

	For $B_\theta\simeq 0$, the polar EM
	force reduces to (Appendix~\ref{app:ftheta_derivation})
	\begin{equation}\label{eq:ftheta_full}
		f_\theta	= -\underbrace{\frac{1}{8\pi r}\partial_\theta(B_r^2+B_\phi^2)}_{\text{pressure gradient}} \, - \,	\underbrace{\frac{\cot\theta}{4\pi r}\,B_\phi^2}_{\text{hoop stress}}
	\end{equation}

	\begin{figure}
		\centering
		\includegraphics[width=0.45\textwidth,clip=false,trim= 0 0 0 0]{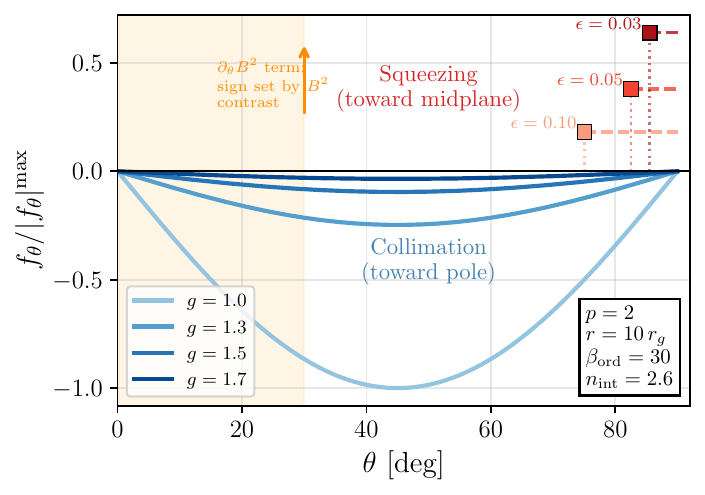}
		\caption{Polar EM force density at $r=10\,r_g$ in an axisymmetric toy setup. Blue curves show $f_\theta$ from Eq.~\eqref{eq:ftheta_wedge} for the
		illustrative field $B_r\propto\cos\theta$, $B_\phi\propto\sin\theta$, normalized to the peak collimation amplitude at $g=1$; for $g<p=2$ the force is poleward ($f_\theta<0$), i.e.\, collimating. Red dashed lines show the interface squeezing $|f_\theta^{\rm sq}|\sim B_r^2/(8\pi r\epsilon)>0$ (Eq.~\eqref{eq:ftheta_squeeze}), directed disk-ward. The orange shaded region is an example of a jet--funnel boundary, where the smooth wedge profiles break down: a $B^2$ transition activates the pressure-gradient term of Eq.~\eqref{eq:ftheta_full}, whose sign is set by the field contrast and can oppose the hoop stress.}
		\label{fig:ftheta_sign}
	\end{figure}
	In the body of the wedge, the magnetic pressure is nearly uniform ($B^2$ is nearly uniform in $\theta$), since $\beta<1$ and thermal support is practically irrelevant. The pressure-gradient term in Eq.~\eqref{eq:ftheta_full} then vanishes, and only the hoop stress survives:
	\begin{equation}\label{eq:ftheta_wedge}
		f_\theta^{\rm wb} \simeq -\frac{\cot\theta}{4\pi r}\,B_\phi^2 \;\leq\; 0
		\qquad (\text{in the wedge body})
	\end{equation}
	directed poleward. This sign is independent of
	$(p,g)$ and follows from $B_\phi^2\geq 0$ and $\cot\theta>0$ for
	$\theta\in(0,\pi/2)$. The wedge body generally produces collimation from the toroidal hoop stress. 

	However, the finite disk thickness and the topology of the ordered field modify this behavior near the disk--wedge interface. A coherent poloidal field threading the disk is antisymmetric across $\theta=\pi/2$ and the toroidal field generated by differential winding reverses with it. The ordered $B^2$ is therefore symmetric about the midplane with a minimum at $\theta=\pi/2$, while the {total} $B^2$ at the midplane is bounded from below by the local MRI turbulence. The higher-$\beta_{\rm g}$ midplane can sustain MRI turbulence, with differential rotation preferentially amplifying toroidal fluctuations. A coherent wedge with $B^2_{\rm wedge}>B^2_{\rm mid}$ requires the ordered wedge field to dominate MRI turbulence in the disk body, $\beta_{\rm g}\lesssim 1/\alpha_{\rm tur}\sim 10$--$30$, the threshold that marks the transport transition of \S\ref{sec:angTransport}.
	
	When this condition holds, $B$ varies from its wedge value to its midplane value across the disk's angular thickness $\Delta\theta\sim\epsilon$, and:
	\begin{equation}\label{eq:ftheta_squeeze}
		f_\theta^{\rm sq}\sim\frac{1}{8\pi r}|\partial_\theta B^2|
		\sim\frac{B_r^2}{8\pi r\epsilon},
	\end{equation}
	where $B_r$, the bent component of the accumulated poloidal flux, dominates the gradient because it varies most sharply across the interface. With Alfv\'enic twist removal, $q\sim\Omega t_{\rm loss}$ and $t_{\rm loss}\sim\ell/v_A$, a midplane density peak favors stronger winding through $v_A\propto\rho^{-1/2}$ at comparable poloidal field strength and loss length $\ell$. We assume that this makes the field increasingly toroidally dominated toward the midplane, where the turbulent toroidal field remains finite while the ordered radial (poloidal) component vanishes by antisymmetry, with the turbulent toroidal pressure approximately compensating the declining ordered toroidal pressure.
    
    The gradient term is then enhanced by $\sim 1/\epsilon$, while the hoop stress is suppressed by $\cot\theta_d\simeq n_{\rm int}\epsilon\ll 1$. The pressure gradient therefore exceeds the hoop stress by $\sim (2q^2n_{\rm int}\epsilon^2)^{-1}$.
	The dynamical significance relative to the disk's support is,
	\begin{equation}\label{eq:ftheta_ratio}
		\frac{f_\theta^{\rm sq}}{P_{\rm mid}/H}
		\sim \frac{1}{(1+q^2)\Pi_{\rm ord}},
	\end{equation}
	In the assumed scenario, the radial-field contrast supplies the squeezing stress. If the ordered field dominates transport ($\beta_{\rm g}\lesssim1/\alpha_{\rm tur}$), squeezing is perturbative for $\Pi_{\rm ord}\gg1$ ($\Psi\ll1$) and becomes dynamically significant when $\Psi\sim1$. The presence of $f_\theta^{\rm sq}$ is consistent with GRMHD simulations in \citet{scepi+24}.

    An analogous pressure-gradient contribution arises at low $\theta$ near the jet--funnel wall. In MAD-like states the funnel is threaded by horizon-anchored BZ flux with $\sigma\gg 1$, while the wedge sits at a much lower magnetization \citep{tchekhovskoy+11,mckinney+12,narayan+22}. 
    If the funnel field dominates, $B^2$ decreases with increasing $\theta$ across the wall, so $\partial_\theta B^2<0$ and the first term in Eq.~\eqref{eq:ftheta_full} gives $f_\theta>0$, directed equatorward and opposing the poleward hoop stress. Collimation of the funnel wall then requires the hoop stress to dominate. Unlike the disk--wedge interface, where antisymmetry fixes both the sign and magnitude of the gradient, the funnel-wall amplitude depends on $B^2_{\rm funnel}-B^2_{\rm wedge}$ and the wall thickness, neither captured by the smooth wedge profiles used in Figure~\ref{fig:ftheta_sign}.

	\subsection{Normalized flux and confinement at $r_c$}
	
\begin{figure}
		\centering
		\includegraphics[width=0.45\textwidth,clip=false,trim= 0 0 0 0]{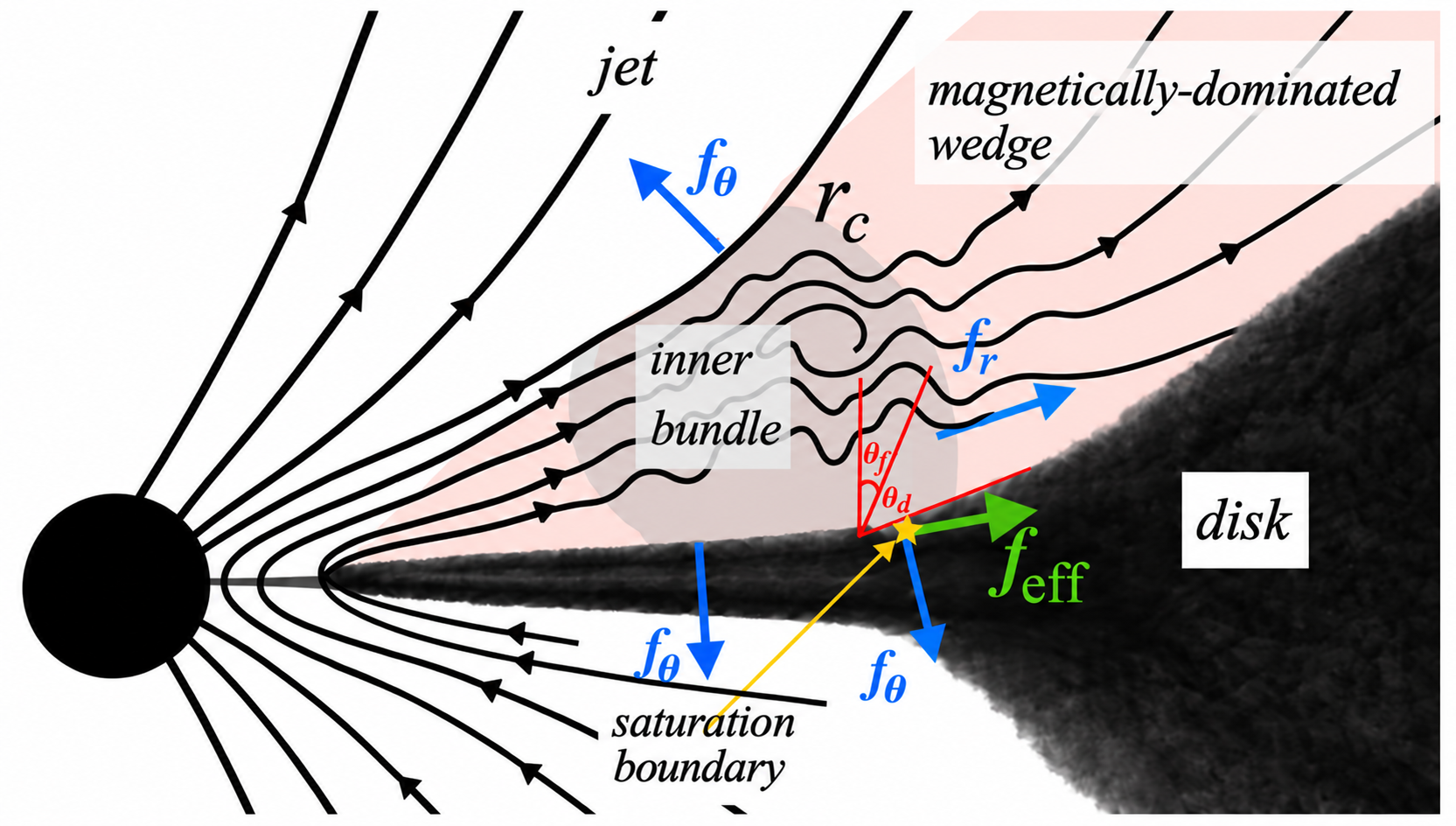}
		\caption{Idealized cartoon of the magnetically dominated wedge,
$\theta_f(r)\leq\theta\leq\theta_d(r)$, and its turbulent,
X-ray-emitting inner flux bundle (gray). Black curves schematically
represent the poloidal magnetic field. Arrows indicate magnetic squeezing, the polar magnetic force density $f_\theta$, and $f_r$. The star marks the saturation boundary at the interface between the disk and the magnetically dominated wedge,
at $r_c\sim10\,r_g$. The broad wedge shown is illustrative.
Narrower wedges and truncated-disk geometries are also possible
provided a saturation region forms with a consistent force balance.
Sizes are not to scale.}
		\label{fig:cartoon}
	\end{figure}

	In the flux-saturated inner bundle near $r\sim r_c$, the accumulated large-scale field is confined at the disk--wedge interface by the inflow inertia and gas pressure of the disk surface, while the wedge above remains strongly magnetized. As flux builds up, the Lorentz force in the bundle becomes comparable to the inflow dynamical pressure, halting further steady inward advance. Outward radial forcing and the inward growth of $P_{\rm B}$ oppose deeper penetration. The interface acts as a magnetically pressurised barrier regulated by intermittent reconnection and interchange (Figure~\ref{fig:cartoon}), so a time-averaged confinement condition holds at $r_c$,
	\begin{equation}\label{eq:confinement}
		\frac{B^2(r_c)}{8\pi} \;\simeq\;
		\xi\,\rho(r_c)\,v_K^2(r_c),
		\qquad
		\xi \;\simeq\; \tfrac{1}{2}\sigma c^2/v_K^2 ,
	\end{equation}
	with
	$v_K(r)=c\sqrt{r_g/r}$ the Keplerian speed, such that $\xi$ measures the magnetic support at the interface. The density $\rho(r_c)$ is evaluated in the magnetically dominated wedge above the disk
	surface. We adopt the GRMHD horizon-flux convention in which the mass flux  through the sphere $r=r_c$ is written as $\dot M_c \;\simeq\; 4\pi r_c^2\,\rho(r_c)\,v_r(r_c)$ with $\rho(r_c)$ the spherically-averaged density at $r_c$ \citep{tchekhovskoy+11,narayan+22}. 
	
	Since the enclosed poloidal flux is approximately constant for
	$r<r_c$, and the bundle's natural flux normalization is the
	hemisphere flux $\Phi_c=2\pi r_c^2\eta_r B(r_c)$, the
	normalized flux is (Appendix~\ref{app:sat_flux_der}):
	\begin{equation}\label{eq:norm_flux}
		\phi =\frac{\Phi_c}{(\dot M_c r_g^2 c)^{1/2}}
		\;\simeq\; 2\sqrt{2}\,\pi\,\eta_r\,\sqrt{\frac{\xi}{\zeta}}
		q_c^{\frac{3}{4(p-g)}}
	\end{equation}
	where $\zeta =  v_r(r_c)/v_K(r_c)$ and
	$\eta_r= (1+q_c^2)^{-1/2}$ accounts for the toroidal winding
	in the total field. The flux $\phi$ is set by the competition
	between magnetic support ($\xi$), mass advection ($\zeta$), and $q_c$.

    \paragraph*{The saturation-boundary pressure ratio.}
At $r_c$, magnetic squeezing compresses the gas, raising its supporting pressure until it balances the magnetic-pressure contrast across the disk--wedge interface. Including radiation, the supporting pressure is $P_{\rm sup}=P_{\rm g}+P_{\rm rad}$.
Retaining the assumed field geometry, $B_r$ carries the dominant vertical gradient and supplies the squeezing force density of Eq.~\eqref{eq:ftheta_squeeze}, so $P_{\rm sup}(r_c)=B_r^2(r_c)/(8\pi)$. The toroidal component adds to the total magnetic pressure, $P_B=B_r^2(1+q_c^2)/(8\pi)$. We assume that MRI turbulence in the higher-$\beta_{\rm g}$ midplane sustains toroidal magnetic pressure through differential winding of radial magnetic fluctuations, approximately offsetting the declining ordered toroidal contribution and leaving a weak toroidal-pressure gradient and negligible toroidal squeezing.
Thus,
\begin{equation}\label{eq:betac_topology}
\Pi_c=\frac{P_{\rm sup}(r_c)}{P_B}
=\frac{1}{1+q_c^2},
\qquad
\beta_{{\rm g},c}
=\frac{1}{\Xi_c(1+q_c^2)},
\end{equation}
where $\Xi_c=1+P_{\rm rad}(r_c)/P_{\rm g}(r_c)$.
The topology therefore fixes the support-to-magnetic pressure ratio $\Pi_c$. The gas plasma beta $\beta_{{\rm g},c}$ additionally depends on the local radiation pressure. In very thin disks, magnetically supported configurations (e.g., $\epsilon \simeq 0.03$), squeezing can be subdominant and the balance in Eq.\eqref{eq:betac_topology} occurs at $\Pi_c = 1$.

	\begin{figure}
		\centering
		\includegraphics[width=0.49\textwidth,clip=false,trim= 0 0 0 0]{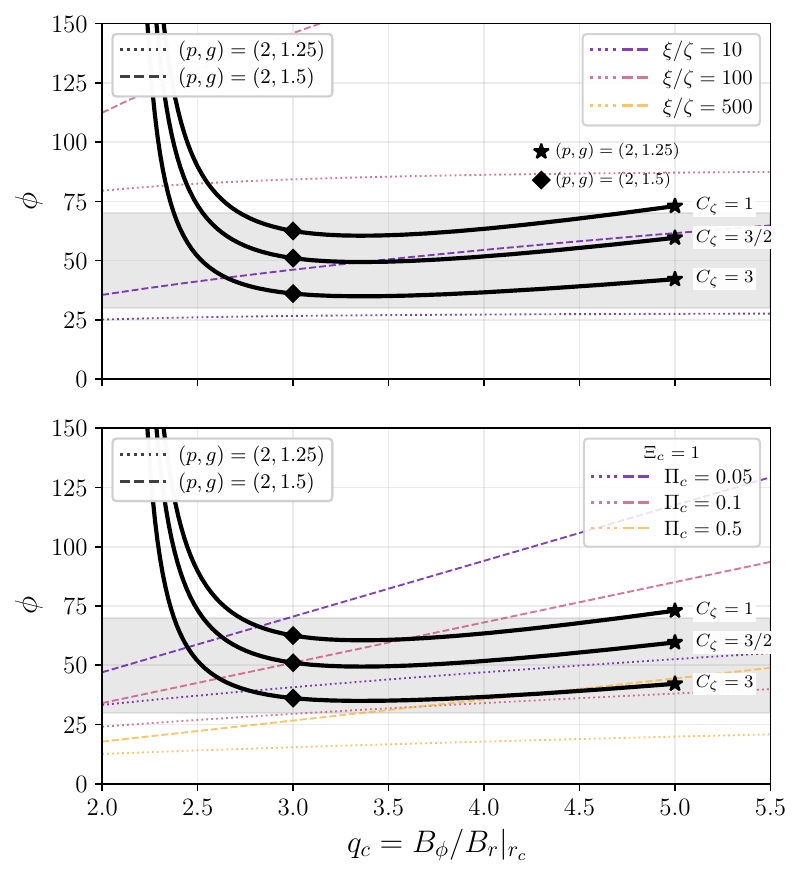}
		\caption{ Normalized flux $\phi$ versus critical twist $q_c=B_\phi/B_r|_{r_c}$. \emph{Upper:} fixed $\xi/\zeta=10,\,100,\,500$ using Eq.~\eqref{eq:norm_flux}. \emph{Lower:} fixed $\Pi_c=5\times10^{-2},\,10^{-1},\,5\times10^{-1}$ using Eq.~\eqref{eq:xi_zeta_betac}. Dotted and dashed colored curves show $(p,g)=(2,1.25)$ and $(p,g)=(2,1.5)$, respectively. The gray band marks $\phi\in[30,70]$, bracketing the GRMHD MAD plateau. Black curves show the topological tracks obtained by varying $g$ at fixed $p=2$ based on Eq.~\eqref{eq:phi_simplified}. The labels indicate the adopted $C_\zeta \sim \mathcal{O}(1)$, which are illustrative rather than a prediction of standard viscous theory. The star and diamond mark the $(p,g)=(2,1.25)$ and $(2,1.5)$ endpoints, respectively. 
		}
		\label{fig:flux_sat}
	\end{figure}

	\paragraph*{Mass advection:}
	The parameter $\zeta$ measures the radial inflow speed at the saturation boundary. Using $\alpha_c$, the ordered Maxwell-stress parameter at the boundary, we get from the thin-disk-like scaling:
	\begin{equation}\label{eq:zeta_c}
		\zeta = \frac{v_r(r_c)}{v_K(r_c)}
		\simeq C_\zeta \frac{\nu_c}{r_c v_K(r_c)}
		\simeq C_\zeta\,\alpha_c\,\epsilon_c^2
	\end{equation}
    where the standard viscous normalization ($v_r=3\nu/(2r)$, $\nu=\alpha_c c_s H$) gives $C_\zeta=3/2$ \citep{pringle81}. At the saturation boundary, however, $\zeta$ need not equal the standard thin-disk value. The boundary layer is a magnetically compressed interface rather than a passive viscous annulus. Ordered magnetic stresses can enhance local angular-momentum loss, raising $C_\zeta$. However, the magnetic barrier at $r_c$ throttles the time-averaged inflow, which could push $C_\zeta$ below $3/2$. We therefore retain the geometric scaling $\zeta\propto\alpha_c\epsilon_c^2$ and bracket the non-thin-disk character of the boundary as $C_\zeta\in[1,3]$ (Figure~\ref{fig:flux_sat}). We use $\alpha_c\simeq 2q_c/[(1+q_c^2)(1+\beta_{{\rm g},c})]$ from Eq.~\eqref{eq:betaLS_threshold_alphaMRI}, with the ordered stress normalized by $\beta_{{\rm g},c}=\Pi_c/\Xi_c=1/[\Xi_c(1+q_c^2)]$ (see Eq.~\eqref{eq:betac_topology}).

	\paragraph*{The ratio $\xi/\zeta$.}
	Adopting $P_{{\rm sup},c}/\rho_c\simeq\Omega_K^2H_c^2$ gives $\xi\simeq\epsilon_c^2/\Pi_c$, where $\epsilon_c=H_c/r_c$.
Since the adopted inflow scaling is
$\zeta\simeq C_\zeta\alpha_c\epsilon_c^2$
(Eq.~\eqref{eq:zeta_c}), the common factor $\epsilon_c^2$
cancels in $\xi/\zeta$, yielding,
		\begin{equation}\label{eq:xi_zeta_betac}
		\frac{\xi}{\zeta}
		\;\simeq\; \frac{(1+q_c^2)(1+\beta_{{\rm g},c})}{2C_{\zeta}\,q_c\,\Pi_c}
		= \frac{(1+q_c^2)\big[(1+q_c^2)+\Xi_c^{-1}\big]}{2C_{\zeta}\,q_c}
	\end{equation}
	using Eq.~\eqref{eq:betac_topology}. The ratio depends on $q_c$, $C_{\zeta}$, and only weakly on the boundary radiation content through $\Xi_c$. It reduces to $(1+q_c^2)(2+q_c^2)/(2C_\zeta q_c)$ in the gas limit ($\Xi_c=1$) and to $(1+q_c^2)^2/(2C_\zeta q_c)$ for $\Xi_c\gg1$. For $q_c=3$–$5$, the two limits differ by approximately $4$–$10\%$ in $\xi/\zeta$, corresponding to approximately $2$–$5\%$ in $\phi$.

	
	\paragraph*{Prediction of the MAD plateau.}
	For $\Pi_c\ll 1$ and $\eta_r\sqrt{1+q_c^2}=1$, 
	Eq.~\eqref{eq:norm_flux} simplifies to (using Eq.~\eqref{eq:betac_topology}):
		\begin{equation}\label{eq:phi_simplified}
		\phi = 2 \pi
		\sqrt{\frac{(1+q_c^2)+\Xi_c^{-1}}{C_\zeta q_c}}
		\,q_c^{\frac{3}{4(p-g)}}
	\end{equation}
	which depends only on the field topology $(p,g)$ through $q_c$ and the
	factor $q_c^{3/[4(p-g)]}=(r_c/r_g)^{3/4}$. For $(p,g)=(2,1.25)$, $q_c = 5.0$, giving $\phi\simeq60$ ($\Xi_c=1$) to $\phi\simeq58$ ($\Xi_c\gg1$) for $C_\zeta=3/2$. Across $(p,g)\in[(2,1.25),(2,1.5)]$ the normalization gives $\phi\simeq 50$--$60$ for $C_\zeta=3/2$, within the MAD range $\phi\simeq30$--$70$ found in GRMHD \citep[e.g.,][]{tchekhovskoy+11}. Since $\phi\propto C_\zeta^{-1/2}$, the order-unity uncertainty in $C_\zeta$ maps to a factor $\lesssim 2$ in $\phi$ (e.g.\ $\phi\simeq36$--$42$ for $C_\zeta=3$), keeping the track inside the plateau. It is worth mentioning that strong squeezing is not required for flux saturation. If magnetic support at smaller thermal $H/r$ reduces the midplane-to-surface density contrast, the resulting weaker enhancement of toroidal dominance toward the midplane could invalidate the assumed compensation of ordered toroidal pressure. If $\Pi_c\simeq1$ replaces Eq.~\eqref{eq:betac_topology}, the same inflow closure gives $\phi\simeq10$--$20$ for the fiducial topologies.

	The near-plateau arises because Eq.~\eqref{eq:betac_topology} ties $\Pi_c$ to $q_c$ in a way that partially offsets the growth of the geometric factor $q_c^{3/[4(p-g)]}\simeq(r_c/r_g)^{3/4}$. The compensation is not exact: as $g$ approaches $p$, $q_c$ decreases only modestly,
	whereas $1/(p-g)$ grows, driving $r_c$ and therefore $\phi$ upward. Figure~\ref{fig:flux_sat} shows $\phi(q_c)$ at fixed $\xi/\zeta$ (upper panel) and fixed $\Pi_c$ (lower panel), with Eq.~\eqref{eq:phi_simplified} overlaid in black with dependence on $C_\zeta$. The colored background curves illustrate how $\phi$ would vary if $\Pi_c$ or $\xi/\zeta$ were treated as free parameters. Values of $\xi/\zeta \sim 10$ and $\Pi_c \sim 0.1$ are favored.
	
	\paragraph*{Estimated $g$-slopes.}
	The slope $g$ is controlled by how rapidly twist is removed. Since $q= B_\phi/B_r\sim\Omega t_{\rm loss}$, while $B_r\propto r^{-p}$ and $B_\phi\propto r^{-g}$ imply $q\propto r^{p-g}$, nearly Keplerian rotation gives
	$p-g\simeq d\ln t_{\rm loss}/d\ln r-3/2$. 
	
	We estimate twist removal by radial Alfv\'enic propagation, $t_{\rm loss}\sim r/v_{A}$, where $v_{A}\simeq |B_r|/\sqrt{4\pi\rho}$ in the nonrelativistic limit. Writing $v_{A}\propto r^{-\Lambda}$ gives $g=p-\Lambda+1/2$. In the saturated bundle, $p=2$, while constant $\dot M_c$ and $v_r\propto v_K$ give $\rho\propto r^{-3/2}$. Thus $v_{A}\propto r^{-5/4}$ and $g=5/4$. Profiles with $1\leq\Lambda<3/2$ give $1<g\leq3/2$. For example, $\rho\propto r^{-2}$ gives $\Lambda=1$ and $g=3/2$.
	
	In the cold-plasma relativistic extension, the radial propagation speed in the local fluid frame is $v_A/c=(1+q^2)^{-1/2}[\sigma/(1+\sigma)]^{1/2}$. For $p=2$ and $\rho\propto r^{-3/2}$, $\sigma\propto(1+q^2)r^{-5/2}$, giving $\Lambda_{\rm rel}=5/4+(1/2)\,d\ln(1+\sigma)/d\ln r$ and $g=p-\Lambda_{\rm rel}+1/2$.
	
	\subsection{Synthesis and connection to MAD phenomenology}
	
	Figure~\ref{fig:cartoon} summarizes the self-regulated coronal wedge picture developed above. The radial extent of the structure is set by the field topology: the twist $q$ approaches the critical value $q_c(p,g)$ where $f_r$ changes sign, defining the saturation radius $r_c$. Inside $r_c$, additional advected flux is preferentially shed, so the enclosed flux saturates and $B_r\propto r^{-2}$. The angular extent, on the other hand, is controlled by the vertical magnetization of the disk atmosphere. In practice, the wedge opens where $\beta<1$, with its boundary set by the local ordered field strength, disk thickness, and mass loading through Eq.~\eqref{eq:ff_wedge_condition}. The critical disk--wedge interface at $r_c$ becomes an important region where radial flux regulation, vertical magnetic squeezing, and flux shedding are coupled most directly. 
	
	These scalings provide a natural interpretation of flux saturation, dissipation, and variability in MAD simulations. In GRMHD, horizon flux saturation is identified by a plateau in the horizon-threading magnetic flux and the onset of magnetically arrested inflow within $\sim 10\,r_g$ \citep[e.g.,][]{narayan+22}. In the present model, the analogous scale is the saturation radius $r_c$, where the accumulated inner flux is confined by the inertia-carrying inflow and excess flux is shed. 
	High-resolution MAD simulations show enhanced dissipation, plasmoid-mediated reconnection, and rapid variability in the inner magnetosphere \citep{ripperda+20,ripperda+22}, which we identify with the flux shedding expected near the saturation layer. The key connection to MAD phenomenology is in the self-regulated balance between flux accumulation, magnetic arrest, and intermittent flux loss near $\sim 10\,r_g$, mostly set by the field topology.

	\paragraph*{Connection between the MAD state and the ordered supply.}
	The wind-fed supply invoked here is most relevant to radiative-mode AGN, which we connect phenomenologically to their environments in \S\ref{sec:observations}. It is not, however, required to form a MAD state under the scalings of \S\ref{sec:dynamics}; ordered flux could instead arise, for example, through dynamo action \citep[e.g.,][]{ide+24,liska+20}. The wind-fed supply in Figure~\ref{fig:roadmap} controls the switch at outer scales, whereas the topology of the resulting quasi-force-free region controls the saturated state, which depends only on a generic supply of ordered flux.
	
	Overall, (i)~the supply magnetization is not predictive, since even $\beta_{\rm w}\sim 10^{2}$ is rescaled during advection by transverse compression and winding,
	$\beta_{\rm g}(r)\propto\beta_{\rm w}(r/R_i)^4/[(1+q^2)f_\Phi^2]$ (Eq.~\eqref{eq:betaLS_from_betaw}), which decreases inward for $\Gamma_{\rm ad}<2$ as $\beta_{\rm g}\propto r^{4-2\Gamma_{\rm ad}}$. The support ratio follows as $\Pi_{\rm ord}=\Xi\beta_{\rm g}$ with $\Xi$ slowly varying, and only the local value near $\gtrsim 10\,r_g$ matters. (ii)~The relevant target for the transport handover is $\beta_{\rm g}\sim10$, not $\beta_{\rm g}\ll1$, as ordered transport overtakes MRI for $\beta_{\rm g}\lesssim2q/[\alpha_{\rm tur}(1+q^2)]-1\sim10$--$30$ (Eq.~\eqref{eq:betaLS_threshold_alphaMRI}), while magnetic domination reaches a substantial fraction of the column as the support ratio $\Pi_{\rm ord}\to\mathcal{O}(\mathrm{few})$ (Eq.~\eqref{eq:squeezing_onset}), with strong compression requiring $\Psi\gtrsim1$. Since $\Pi_{\rm ord}=\Xi\beta_{\rm g}\ge\beta_{\rm g}$, the squeezing stage lies deeper in. The threshold is also twist-weighted, so winding with $q\gtrsim1$ both facilitates the handover and sets the saturation radius through $r_c/r_g\simeq q_c^{1/(p-g)}$ (Eq.~\eqref{eq:qc_rc}).

	\section{Observable consequences}\label{sec:observations}
	
	\subsection{The flux-saturated inner bundle as an X-ray corona}
	
	\subsubsection{Flux saturation and general coronal constraints}
	
	The flux-saturated inner bundle (with characteristic scale $r_c$) is a natural candidate for the AGN X-ray corona. The saturation region is intrinsically compact ($r_c\sim 10\,r_g$) and time dependent, with variability ranging from Alfv\'enic turbulent times to longer accretion-driven dynamical times. Its low density and magnetic dominance ($\beta \ll 1$ and modest to large magnetization $\sigma \sim 1$) make magnetic dissipation and particle energization collisionless at kinetic scales: the Coulomb mean free path greatly exceeds the skin depth, and ion relaxation and electron--ion exchange are slower than the light-crossing time, though electron--electron scattering may still relax a weakly collisional thermal core, motivating particle-in-cell treatments \citep[e.g.,][]{kunz+14b,sironi+15b}. 
    
    By impeding the inflow, the saturation region provides a natural site for converting the accretion power associated with $\dot M_{\rm tot}$ into magnetic energy, turbulent motions, and coronal heating.
    Turbulent dissipation can then maintain hot electrons with nonthermal tails and modest optical depth, enabling inverse-Compton upscattering of disk UV seed photons into a hard X-ray power law with a thermal cutoff, while reconnection at current-sheet and flux-shedding boundaries can supply injection \citep[e.g.,][]{ripperda+20,ripperda+22}. Finally, the small, luminous geometry places the region in the high-compactness regime where radiative cooling and pair creation regulate temperature and optical depth \citep[e.g.,][]{fabian+15,fabian+17,ricci+18}, and its location above the inner disk in an extended, magnetically supported disk–funnel interface naturally produces strong illumination required for relativistic reflection and reverberation lags. In this geometry the Fe~K$\alpha$ line can be emitted as hard coronal X-rays irradiate the optically thick disk and produce fluorescence together with the reflected continuum \citep{george+91}, with the wedge geometry setting the illumination pattern. We can then derive scaling relations for the global coronal properties that shape the observed X-ray phenomenology, and connect them to the compactness, heating, and confinement conditions explored by kinetic models.

	\subsubsection{Flux shedding as the driver of coronal turbulence}\label{sec:turbulence}
	
	Inside the flux-saturated inner bundle, the mean enclosed poloidal flux is approximately steady. Interchange-driven flux shedding redistributes the field and current density, triggering Alfv\'enic readjustment and the formation of eddies and current sheets. Global GRMHD simulations of MAD states confirm the presence of time-dependent inner flows with interchange-driven eruptions, current-sheet formation, and turbulent stresses \citep[e.g.,][]{narayan+03,igumenshchev08,tchekhovskoy+11,mckinney+12,manikantan+24,nalewajko+24}. This intermittently forced turbulence supplies coronal energy that heats and accelerates electrons, which upscatter disk photons into X-rays. Radiative PIC studies support this picture, showing efficient electron heating, sustained nonthermal tails, and hard Comptonized X-ray spectra whose variability reflects the intermittency of turbulent structures \citep[e.g.,][]{groselj+24,groselj+26,mbarek+26a}. Reconnection-based models can in principle also generate hard X-rays, but in the present picture the global power and variability are more naturally set by turbulence driven by flux balance, with reconnection remaining potentially important as an injection channel at boundaries \citep[e.g.,][]{mbarek+22,mbarek+24,ripperda+22}. Turbulence is expected to be most vigorous near $r\sim r_c$ where flux eruptions and interchange are triggered \citep[e.g.,][]{tchekhovskoy+11,mckinney+12,wang+25}.

	\subsubsection{Threshold coronal magnetization}

    Two-temperature disk simulations develop hot, magnetically dominated regions above the disk \citep{bambic+24a}, while radiative kinetic models find weak ion–electron Coulomb coupling for $\sigma \gtrsim 0.04$ under representative coronal conditions \citep{groselj+26}. These results motivate a magnetized, weakly collisional corona, but do not by themselves establish whether magnetic stresses control its bulk dynamics. To assess their dynamical importance in the proposed model, we compare the polar and radial EM force densities in the wedge body with the gravitational force density (Appendix~\ref{app:force_ratio_gr}).

	In the wedge body, the main polar force is the toroidal hoop stress $f_\theta^{\rm wb} \simeq -\frac{\cot\theta}{4\pi r}B_\phi^2$. With $B_\phi^2=B^2q^2/(1+q^2)$, and $\sigma= B^2/(4\pi\rho c^2)$, and normalizing by the Kerr-corrected ZAMO gravity through $\mathcal{R}_{\rm GR}(r,a_\ast)= c^2/(r\,g_r^{\rm Kerr})$, we get:
	\begin{equation}
		\label{eq:gr_sigma_theta}
		\frac{|f_\theta^{\rm wb}|}{|\rho g_r|}
		\simeq
		\sigma\,\mathcal{R}_{\rm GR}(r,a_\ast)
		\frac{q^2}{1+q^2}\cot\theta .
	\end{equation}
	Thus the wedge-body polar magnetic force density exceeds the gravitational force density for
	\begin{equation}
		\sigma
		\gtrsim
		\frac{1+q^2}
		{\mathcal{R}_{\rm GR}(r,a_\ast)\,q^2\cot\theta}.
	\end{equation}
	For $q\simeq2$, $\theta\simeq45^\circ$, and $r\simeq10\,r_g$, the Kerr ZAMO normalization gives $\mathcal{R}_{\rm GR}\simeq9$ across the allowed spin range, yielding $\sigma\gtrsim0.15$, comparable to the regime where PIC simulations find nonthermal power-law tails to arise readily and harden with increasing $\sigma$ \citep[e.g.,][]{zhdankin+18,comisso+18,mbarek+24}.
	
\begin{figure}
		\centering
		\includegraphics[width=0.45\textwidth,clip=false,trim= 0 0 0 0]{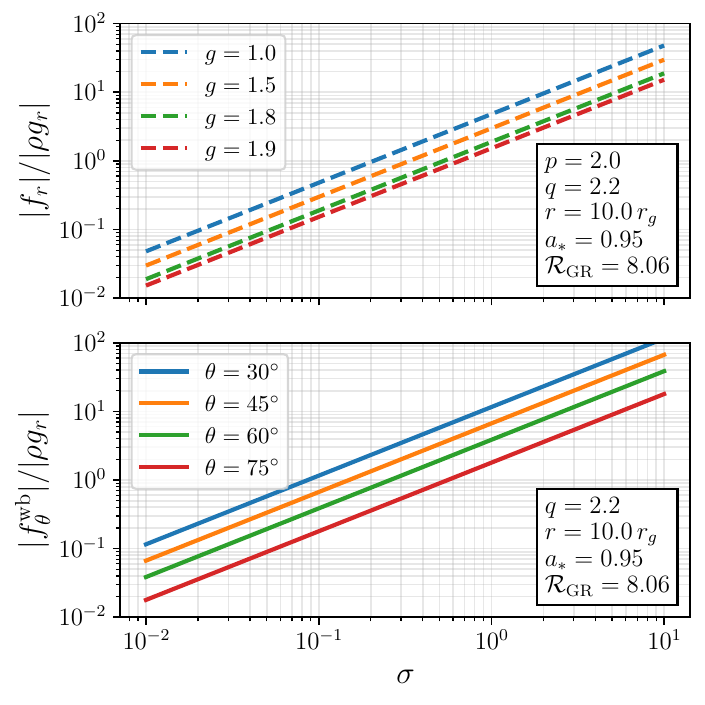}
		\caption{Electromagnetic to gravitational force ratio at $r=10\,r_g$ with Kerr corrections for $a_\ast=0.95$. Top: radial estimate, $|f_r|/|\rho g_r|$, for several toroidal slopes $g$ at fixed $p=2$ and $q=2.2$. Bottom: wedge-body hoop-stress estimate, $|f_\theta^{\rm wb}|/|\rho g_r|\simeq \sigma\mathcal{R}_{\rm GR}[q^2/(1+q^2)]\cot\theta$, for several polar angles. The enhanced squeezing force is not shown because it applies only to the thin disk--wedge transition layer, not to the coronal wedge body.}
		\label{fig:em_vs_gravity_ratio}
	\end{figure}

	The $f_r$ diagnostic provides a complementary clue. For $B_r\propto r^{-p}$ and $B_\phi\propto r^{-g}$,
	\begin{equation}
		\label{eq:gr_sigma_radial}
		\frac{f_r}{|\rho g_r|}
		\simeq
		\frac{\sigma\,\mathcal{R}_{\rm GR}}{1+q^2}\,\mathcal{F}(q),
	\end{equation}
	Radial EM stresses become dynamically important for
	\begin{equation}
		\sigma \gtrsim
		\frac{1+q^2}{\mathcal{R}_{\rm GR}\,|\mathcal{F}(q)|}.
	\end{equation}
    Both ratios are shown in Figure~\ref{fig:em_vs_gravity_ratio}: the radial estimate $|f_r|/|\rho g_r|$ (top) for several toroidal slopes $g$, and the wedge-body hoop-stress estimate $|f^{\rm wb}_\theta|/|\rho g_r|$ (bottom) for several polar angles. For the considered field geometries, EM force densities become comparable to or exceed the gravitational force density at $\sigma \gtrsim 0.1$, allowing an appreciably magnetized buffer zone to control the dynamics.
	
	\subsubsection{Characteristic and maximum coronal magnetization}
	
	A characteristic coronal magnetization can be estimated by combining the magnetic-energy budget with the mass flux loaded into the X-ray-emitting region. We write this mass flux as $\dot M_c=f_{\dot M}\dot M_{\rm Edd}$, with $\dot M_{\rm Edd}=L_{\rm Edd}/(\eta_{\rm rad}c^2)$, where $\eta_{\rm rad}$ is the radiative efficiency of the underlying accretion flow. Following \citet{groselj+24}, we write $U_B=B^2/(8\pi)=\kappa_{\rm x}U_X$, where $U_B$ and $U_X$ are the magnetic and X-ray photon energy densities, respectively, and $\kappa_{\rm x}$ is their ratio. As discussed in Appendix~\ref{app:thermal_compton}, $\kappa_{\rm x}=\mathcal{O}(1)$ when the Comptonized component dominates and photon escape and turbulent dissipation occur on comparable timescales. Using the spherical estimate $U_X\simeq L_X/(4\pi r_c^2c)$ and mass conservation, $\dot M_c=4\pi r_c^2\rho v_r$, the magnetization $\sigma=B^2/(4\pi\rho c^2)$ becomes:
	\begin{equation}\label{eq:sigma_Mdot}
		\sigma \simeq \frac{2 \eta_{\rm rad} \kappa_{\rm x}}{f_{\dot M}}
		\left(\frac{v_r}{c}\right)
		\left(\frac{L_X}{L_{\rm Edd}}\right).
	\end{equation}
	
	The same confinement condition fixes the coronal magnetization  directly. With $\rho_c$ from $\dot{M}_c$ and  $B^2 = 8\pi\xi\rho_c v_K^2$,
	\begin{equation}\label{eq:sigma_conf}
		\sigma=\frac{B^2(r_c)}{4\pi\rho_c c^2}
		=2\xi\,\frac{v_K^2}{c^2}
		=\frac{\phi^2}{4\pi^2\eta_r^2}
		\left(\frac{r_c}{r_g}\right)^{-3/2}\frac{v_K\,v_r}{c^2}.
	\end{equation}
	Adopting 
	$v_r\le v_K$, we get:
	\begin{equation}\label{eq:sigma_bound}
		\sigma \;\le\; \frac{\phi^2}{4\pi^2\eta_r^2}
		\left(\frac{r_c}{r_g}\right)^{-5/2}
		=\frac{\phi^2(1+q_c^2)}{4\pi^2}
		\left(\frac{r_c}{r_g}\right)^{-5/2}.
	\end{equation}
	For $\phi=50$, and $(p,g)=(2,1.5)$ (so $q_c=3$), the  bound gives $\sigma \le 2.5$. For $(p,g)=(2,1.25)$ ($q_c=5$), it gives $\sigma\le 8$. These bounds show that $\sigma \sim 0.1-1$ is allowed if the coronal inflow is weakly loaded or dynamically fast, $v_r \sim (v_K/10)$--$v_K$. Setting $\sigma\sim 1$ in Eq.~\eqref{eq:sigma_Mdot} gives $f_{\dot M}(c/v_r)\sim 2 \eta_{\rm rad}\kappa_{\rm x}L_X/L_{\rm Edd}$, or $f_{\dot M}\sim 2\times 10^{-5}$--$2\times 10^{-3}$ for $\kappa_{\rm x}\sim1$, $v_r/c\sim10^{-2}$--$10^{-1}$, $\eta_{\rm rad} = 0.1$, and $L_X/L_{\rm Edd}\sim10^{-2}$--$0.1$.

	\subsubsection{The X-ray luminosity of the corona}
	We now estimate the X-ray luminosity powered by the same confined  magnetic bundle. Combining the confinement condition 
	with the quasi-spherical mass-conservation relation for $\dot{M}_c$ gives:
	\begin{align}\label{eq:Ub}
		\frac{B^2(r_c)}{8\pi}
		\;&\simeq\;
		\frac{\xi}{\zeta}\,\frac{\dot M_c\,v_K(r_c)}{4\pi\,r_c^2} \nonumber \\
		\;&=\;
		\frac{\phi^2}{32\pi^3\,\eta_r^2}\left(\frac{r_c}{r_g}\right)^{-3/2}
		\frac{\dot M_c\,v_K(r_c)}{r_c^2},
	\end{align}
	where the second equality uses  $\xi/\zeta $ from  Eq.~\eqref{eq:norm_flux}.
	
	The X-ray-emitting inner bundle or corona subtends a solid angle $ \Delta\Omega_X=4\pi(\cos\theta_f-\cos\theta_X),$ where $\theta_f \le \theta_X\le\theta_d $. Here, $\theta_f$ is the funnel boundary, $\theta_d$ is the $\beta=1$ wedge boundary from \S\ref{sec:ang_extent} (Eq.~\eqref{eq:ff_wedge_condition}), and $\theta_X$ parameterizes the outer edge of the hard-X-ray-emitting portion. Writing $\mathcal{A}_X = \cos\theta_f-\cos\theta_X$, we get $L_X\simeq4\pi\mathcal A_Xr_c^2cU_X$ for the wedge geometry. The jet momentum Eq.~\eqref{eq:Ub} gives:
	\begin{equation}\label{eq:LX_phi}
		L_X \;\simeq\; \frac{\mathcal{A}_X\,\phi^2}
		{8\pi^2\,\kappa_{\rm x}\,\eta_r^2}
		\left(\frac{r_c}{r_g}\right)^{-3/2}
		\dot M_c\,v_K(r_c)\,c .
	\end{equation}
	Using $v_K(r_c)/c=(r_c/r_g)^{-1/2}$ and 
	$\eta_r^2=(1+q_c^2)^{-1}$, we get:
	\begin{equation}\label{eq:LX_Mc}
		\frac{L_X}{\dot M_c c^2}
		\;\simeq\;
		\frac{\mathcal{A}_X\,\phi^2(1+q_c^2)}{8\pi^2\kappa_{\rm x}}
		\left(\frac{r_c}{r_g}\right)^{-2}
	\end{equation}

	The $(r_c/r_g)^{-2}$ factor is not a Novikov--Thorne surface-flux profile, which describes local dissipation in an optically thick thermal disk. At fixed $(\phi,q_c,\kappa_{\rm x},\mathcal{A}_X)$, confinement gives $U_B\propto r_c^{-4}$ and the emitting area contributes $r_c^2$, yielding $L_X\propto r_c^{-2}$. Imposing $\phi\propto r_c^{3/4}$ from this model instead gives $L_X/(\dot M_c c^2)\propto r_c^{-1/2}$. This follows from this paper's coronal magnetic geometry, not the disk dissipation law.

	In absolute units,
	\begin{align}\label{eq:LX_numeric}
		\frac{L_X}{10^{42}\,{\rm erg\,s^{-1}}}
		\;\simeq\;\;& 
		\left(\frac{\mathcal{A}_X}{\cos\theta_f-\cos 45^\circ}\right)
		\kappa_{\rm x}^{-1}
		\left(\frac{1+q_c^2}{10}\right)
		\left(\frac{\phi}{50}\right)^2 \nonumber\\
		&\quad\times
		\left(\frac{r_c}{10\,r_g}\right)^{-2}
		\left(\frac{\dot M_c}{2\times 10^{-5}\,M_\odot\,{\rm yr^{-1}}}\right)
	\end{align}
	At $\phi\sim 50$, $r_c\sim 10\,r_g$, and $\theta_X\sim 45^\circ$, an X-ray luminosity of $10^{42}\,{\rm erg\,s^{-1}}$ therefore requires  $\dot M_c\sim 2\times 10^{-5}\,M_\odot\,{\rm yr^{-1}}$. In this form  $L_X$ depends on field topology $(\phi, q_c, r_c/r_g)$ and the absolute coronal mass flux, with no explicit dependence on $M_\bullet$.
	
	Comparing the coronal loading to the total accretion rate inferred 
	from the bolometric luminosity, $\dot M_{\rm tot}=L_{\rm bol}/(\eta_{\rm rad}c^2)$, gives
	\begin{equation}\label{eq:Mc_over_Mtot}
		\frac{\dot M_c}{\dot M_{\rm tot}}
		\;\simeq\; 10^{-2}
		\left(\frac{\dot M_c}{2\times 10^{-5}\,M_\odot\,{\rm yr^{-1}}}\right)
		\left(\frac{\eta_{\rm rad}}{0.1}\right)
		\left(\frac{L_{\rm bol}}{10^{43}\,{\rm erg\,s^{-1}}}\right)^{-1}
	\end{equation}
	Equivalently, because Eq.~\eqref{eq:LX_Mc} gives $\dot M_c\propto L_X$ 
	at fixed topology, the loaded fraction scales as 
	$\dot M_c/\dot M_{\rm tot}\propto L_X/L_{\rm bol}$. For 
	$L_X/L_{\rm bol}\sim 0.01$--$0.1$, only a small fraction of 
	the bulk inflow needs to enter the magnetized X-ray-emitting phase.
	
	Finally, the coronal loading $\dot M_c\sim2\times10^{-5}\,
	M_\odot\,{\rm yr^{-1}}$ is physically undemanding and can be much larger. Through
	$\dot M_c/\dot M_{\rm tot}\sim10^{-2}$,	it implies a total inflow $\dot M_{\rm tot}\sim10^{-3}\,	M_\odot\,{\rm yr^{-1}}$. This can be supplied by capturing only a small fraction $f_{\rm cap}$ of a circumnuclear starburst wind, $\dot M_{\rm tot}\sim f_{\rm cap}\mathcal{M}{\rm SFR}$, where $\mathcal{M}$ is the wind mass-loading factor. For a nuclear star forming region with ${\rm SFR}\sim10\,M_\odot\,{\rm yr^{-1}}$ and a launch region $R_{\rm w}\sim10\,{\rm pc}$, even $f_{\rm cap}\sim10^{-5}$--$10^{-4}$ gives
	$\dot M_{\rm tot}\sim10^{-4}$--$10^{-3}\mathcal{M}\,
	M_\odot\,{\rm yr^{-1}}$, with a fiducial
	$\mathcal{M}\sim1$--a few\footnote{\citet{strickland+09} find mildly mass-loaded wind-fluid solutions for M82 with $\mathcal{M}\simeq1.0$--$2.8$, while broader hot-wind constraints favor $\dot M_{\rm hot}/{\rm SFR}\lesssim1$ for high-SFR systems unless the wind becomes strongly radiative \citep{zhang+14,heckman+17}.}. This is larger than the Galactic-Center	Bondi inflow by roughly two orders of magnitude ($\sim10^{-5}\,M_\odot\,{\rm yr^{-1}}$; \citealt{quataert+99,baganoff+03,quataert04}), as expected for a nucleus with substantially higher star-formation activity.

	\subsubsection{$\lx/\lbol$ is independent of black-hole mass}
	
	The same cancellation can be expressed in terms of the X-ray fraction. Dividing Eq.~\eqref{eq:LX_Mc} by $L_{\rm bol}=\eta_{\rm rad}\dot M_{\rm tot}c^2$,
	\begin{align}\label{eq:LX_Lbol}
		\frac{L_X}{L_{\rm bol}}
		\;\simeq\;\;&
		0.1
		\left(\frac{\mathcal{A}_X}{\cos \theta_f-\cos 45^\circ}\right)
		\kappa_{\rm x}^{-1}
		\left(\frac{1+q_c^2}{10}\right)
		\left(\frac{\phi}{50}\right)^2 \nonumber\\
		&\quad\times
		\left(\frac{\eta_{\rm rad}}{0.1}\right)^{-1}
		\left(\frac{r_c}{10\,r_g}\right)^{-2}
		\left(\frac{\dot M_c/\dot M_{\rm tot}}{10^{-2}}\right) 
	\end{align}
	At fixed magnetic-bundle topology, the X-ray fraction is controlled  by the loaded fraction $\dot M_c/\dot M_{\rm tot}$ of the bulk  inflow that enters the magnetized coronal phase, not by $M_\bullet$. For the fiducial parameters, loading 
	$\dot M_c/\dot M_{\rm tot}\simeq 10^{-2}$ gives 
	$L_X/L_{\rm bol}\simeq 0.1$, comparable to the X-ray fraction of 
	radiative-mode AGN.
	
	This mass cancellation is observationally important. An X-ray luminosity function alone does not uniquely determine the underlying SMBH mass function, because the mapping from $L_X$ to $M_\bullet$ also depends on the Eddington ratio. In particular, $\log\lambda_{\rm E}=\log(L_{\rm bol}/{\rm erg\,s^{-1}})
	-\log(M_\bullet/M_\odot)-38.1$ \citep{ananna+22}, while a commonly used bolometric correction gives $L_X(2$--$10\,{\rm keV})\simeq L_{\rm bol}/20$ \citep{marconi+04}. The same observed X-ray luminosity can therefore arise from a lower-mass black hole accreting at high $\lambda_{\rm E}$ or from a more massive black hole accreting at low $\lambda_{\rm E}$.

	\subsubsection{Kinetic competition: heating, escape, and radiative cooling}
	
	Once EM stresses dominate, particles can radiate before being swallowed by the black hole. Appendix~\ref{app:thermal_compton} discusses the canonical picture where inverse-Compton losses in a UV-brightened compact region lead to $\lambda_{\rm cool}/r_c\ll 1$ when the compactness is $\ell\gtrsim 10$ at the Lorentz factors needed to upscatter UV photons into the X-ray band, naturally placing the system in the efficient, unsaturated thermal Comptonization regime \citep{rybicki+79,sunyaev+80,lightman+87,haardt+91}. Observationally inferred coronal compactness values in AGN are typically at least order unity, consistent with rapid radiative cooling and strong Comptonization \citep[e.g.,][]{fabian+15,fabian+17,ricci+18}. Moreover, in such magnetized, turbulent environments, hadrons should naturally develop a nonthermal population \citep{mbarek+24}, with direct observational consequences \citep[e.g.,][for neutrinos]{ICECUBE25-Xdiffuse}.

	\subsubsection{Coronal magnetic field}
	The confinement condition in Eq.~\eqref{eq:confinement} can fix $B(r_c)$, to have a consistent input to the dissipation and pair-loading estimates below. Combining Eq.~\eqref{eq:confinement} with quasi-spherical mass conservation $\rho(r_c)=\dot M_c/(4\pi r_c^2 \zeta v_K)$ gives $B^2(r_c)=(2\xi/\zeta)\dot M_c v_K(r_c)/r_c^2$, and inserting
	$\xi/\zeta=\phi^2/[8\pi^2\eta_r^2(r_c/r_g)^{3/2}]$ from
	Eq.~\eqref{eq:norm_flux} yields a field set by the saturation
	topology, $\dot M_c$, and $M_\bullet$:
	\begin{equation}\label{eq:Bc}
		B^2(r_c)\;\simeq\;\frac{\phi^2(1+q_c^2)}{4\pi^2}\,
		\frac{\dot M_c\,c}{r_g^2}\left(\frac{r_c}{r_g}\right)^{-4}
	\end{equation}
	Normalizing based on our fiducial values,
	\begin{align}\label{eq:Bc_numeric}
		B(r_c)\;\simeq\;\;& 10^{3}\,{\rm G}\,
		\left(\frac{\phi}{50}\right)
		\left(\frac{1+q_c^2}{10}\right)^{1/2}
		\left(\frac{r_c}{10\,r_g}\right)^{-2} \nonumber\\
		&\times
		\left(\frac{\dot M_c}{2\times10^{-5}\,M_\odot\,{\rm yr^{-1}}}\right)^{1/2}
		\left(\frac{M_\bullet}{10^{7}\,M_\odot}\right)^{-1}
	\end{align}
	so large $B(r_c)$-values reaching $\sim10^{3}\,$G follow from $\phi\simeq50$, $r_c\simeq10\,r_g$, and from $L_X\propto\dot M_c$ (Eq.~\eqref{eq:LX_Mc}) based on $\kappa_{\rm x} \sim 1$.

A field $B\sim10^3\,{\rm G}$ makes synchrotron radiatively important without making it the main source of escaping X-rays. Even when $U_B/U_{\rm ph}>1$, synchrotron self-absorption can recycle the emitted power into the plasma, so the optically thin Thomson-limit ratio $P_{\rm syn}/P_{\rm IC}=U_B/U_{\rm ph}$ does not determine the escaping
spectrum. Mildly relativistic thermal electrons emit synchrotron at radio--mm frequencies, supplying seed photons and promoting thermalization through the synchrotron boiler \citep{ghisellini+88}. Repeated Compton
scattering of disk UV and internal synchrotron photons can produce the observed X-ray continuum \citep{sunyaev+80,haardt+91,fabian+15}. Radiative PIC simulations of turbulent pair coronae in X-ray binaries
demonstrate this internal seed channel, with turbulent bulk Comptonization producing hard-state-like spectra and stronger disk illumination producing soft-state-like spectra \citep{nattila24}. Thus, Comptonization need not
rely solely on thermal electron motions or external seed photons.

\subsubsection{Variability timescales}\label{sec:variability}
	
The shortest expected variability timescale is the light crossing time, $\simeq r_c/c$. Longer variability should arise from two additional effects: (i) magnetic adjustment of the coronal structure, and (ii) accretion-powered changes to the saturated magnetic flux. In particular, the plasma cannot coordinate a global magnetic reconfiguration faster than $r_c/v_A$, where $v_A/c=\sqrt{\sigma/(1+\sigma)}$. In a turbulent corona, such Alfv\'enic rearrangements may be common \citep{mbarek+26a}. For moderately magnetized coronae with, e.g., $\sigma\sim 0.1$, one finds $v_A\sim c/3$, making this a potentially observable variability timescale. Detecting such fast Alfv\'enic variability would be especially valuable because it directly constrains $v_A$ and thus the bulk magnetization $\sigma$. Combined with independent constraints on $\sigma_e$ from compactness, this would provide insight into the proton to pair ratio of the plasma.

These are a hierarchy, not competing flare estimates: $t_{\rm lc}\sim r_c/c$, $t_A\sim r_c/v_A$, and the local dynamical time govern rapid variability of flux already stored near $r_c$, whereas $t_{\rm in}\sim[\alpha(H/r)^2\Omega_K]^{-1}\gg t_A$ governs slow resupply and recurrence.
	
Accretion powered variability can also arise when additional ordered magnetic flux is advected into the inner $\sim 10\, r_g$, enhancing field winding and concentrating toroidal stresses toward small radii. We have $r_c\propto q_c^{1/(p-g)}$ and $\phi\propto q_c^{3/[4(p-g)]}$, so that $\phi\propto r_c^{3/4}$. A larger $r_c$ mainly reflects a geometric expansion of the saturation region and does not by itself imply a higher local energy density at $r\sim r_c$. If the ordered flux $\Phi$ is spread over a larger cross section, then $B(r_c)\sim \Phi/(\pi r_c^2)$ decreases, so $U_B\propto B^2$ is diluted unless $\Phi$ grows even faster. Whether such episodes appear as bright X-ray flares is therefore governed less by the increase in $r_c$ itself than by the radiative efficiency, which is more directly tracked by the compactness $\ell_X\propto L_X/r_c$.
	
This picture is consistent with the reflection-based inference for I~Zw~1 \citep{wilkins+22},  where the effective coronal extent grows from $\sim 7\,r_g$ before the flare to $\sim 18\,r_g$ during the flare, while the soft band flux peaks at $\sim 2.5$ times the pre-flare level. In our scalings, the size increase ($\sim 18/7\simeq 2.6$) together with a comparable luminosity increase implies $\ell_{X,\rm fl}/\ell_{X,\rm pre}\approx 2.5/2.6\simeq 1$, so the corona can brighten even as it expands. Since $\phi\propto r_c^{3/4}$, the same expansion corresponds to an order unity change in flux, $\phi_{\rm fl}/\phi_{\rm pre}\sim (18/7)^{3/4}\approx 2$. Thus a corona that is both larger and brighter can be read as rapid redistribution and expansion of flux accumulated earlier by the slower inflow, not a uniform rise in photon or magnetic energy density.

\subsubsection{Hadron-driven pair loading and compactness regulation}
	
The intense X-ray output, quantified by compactness $\ell_X= L_X\sigma_T/(r_c m_e c^3)$, arises primarily from inverse-Compton scattering of UV seed photons. In the subset of initially ion--electron coronae that develop a sufficiently hard nonthermal proton tail, Bethe--Heitler and photomeson interactions can then inject secondary pairs.

	\paragraph*{\bf{Bethe--Heitler (BHe) pair injection.}}
	With effective cross section $\hat{\sigma}_{\rm BHe}\sim 10^{-27}\,{\rm cm^2}$ and inelasticity $\hat{\xi}_{\rm BHe}\sim 2m_e/m_p\simeq 10^{-3}$ \citep{murase+10b}, the BHe threshold is $\gamma_{p,{\rm BHe}}\sim m_e c^2/\varepsilon\sim 5\times10^2(\varepsilon_X/1\,{\rm keV})^{-1}$, with $\varepsilon$ the photon energy. Writing the X-ray photon number density as $n_\gamma= L_X/(4\pi r_c^2 c\,\varepsilon_X)$, the proton interaction time against this field, $t_{\rm BHe,inj}=(n_\gamma\hat\sigma_{\rm BHe}c)^{-1}$, normalized to the light-crossing time is:
	\begin{equation}
		\frac{t_{\rm BHe,inj}}{r_c/c}
		\simeq 4\pi\,\hat\sigma_{\rm BHe}^{-1}\sigma_T\,
		\frac{\varepsilon_X}{m_ec^2}\,\ell_X^{-1}
		\simeq 16\,\ell_X^{-1}\left(\frac{\varepsilon_X}{1\,{\rm keV}}\right),
	\end{equation}
	where the coefficient $4\pi(\sigma_T/\hat\sigma_{\rm BHe})(\varepsilon_X/m_ec^2)\simeq 16$ for $\varepsilon_X=1\,{\rm keV}$. Thus $\ell_X\sim 10$ gives $t_{\rm BHe,inj}\sim 1.6\,r_c/c$ for protons above threshold, comparable to the light-crossing time. Only high-$\gamma$ protons with $\gamma_p\gtrsim\gamma_{p,{\rm BHe}}$ participate. For a nonthermal proton population, we normalize the available nonthermal power at the characteristic energy selected by turbulent acceleration \citep{mbarek+26a}. If acceleration is efficient at the BHe threshold, $\gamma_{p,{\rm BHe}}$, then the power available for pair injection can be estimated by extrapolating the power per logarithmic interval from $\gamma_p\sim\sigma_p \simeq \sigma$ to $\gamma_p\gtrsim\gamma_{p,{\rm BHe}}$ \footnote{This is a normalization prescription for systems with $\sigma\sim 1$, not an assumption that the full proton distribution is a $dN/d\gamma_p\propto\gamma_p^{-2}$ power law. Numerically it is equivalent to taking comparable power per logarithmic interval between the proton magnetization $\gamma_p\sim\sigma_p$ and $\gamma_p\gtrsim\gamma_{p,{\rm BHe}}$, but the spectral shape is very likely different \citep{mbarek+26a,mbarek26a}.}. We can write $n_{p,>\gamma_{\rm BHe}}/n_p
	\simeq \gamma_{\min}/\gamma_{p,{\rm BHe}}
	\simeq 2\times 10^{-3}\sigma\left(\varepsilon_X/1\,{\rm keV}\right)$. 
	The pair yield per total proton density over residence time $t_{\rm esc}$ is:
	\begin{equation}
		\frac{n_{\pm}}{n_p}
		\sim \frac{n_{p,>\gamma_{\rm BHe}}}{n_p}\,\frac{t_{\rm esc}}{t_{\rm BHe,inj}}
		\sim 1.3\times 10^{-4}\,\sigma\left(c\,t_{\rm esc}/r_c\right)\ell_X
		\label{eq:bh_pair_multiplicity_corrected}
	\end{equation}
	Using the diffusive escape from Eq.~3 in \citet{mbarek+24}, and for $r_c\sim 10\,r_g$, we get:
	\begin{equation}
		\frac{n_{\pm}}{n_p}\ \sim\ 0.13\,\ell_X\sigma
		\left(\frac{\gamma_p}{5\times 10^{2}}\right)^{-1/3}
		\left(\frac{B_c}{10^3\,{\rm G}}\right)^{1/3}
		\left(\frac{M_\bullet}{10^7\,M_\odot}\right)^{1/3}
		\label{eq:bh_pair_multiplicity_diffusive_corrected_gamma}
	\end{equation}
	Pair domination by this BHe channel ($n_\pm/n_p\gtrsim1$) requires a hard proton tail beyond threshold and nominally $\ell_X\sigma\gtrsim 8$, which is an additional, conditional route to the pair-rich states of classical pair balance \citep{svensson84}. Other assumptions, albeit in the standard and normal evolution (SANE) regime, give much smaller fractions \citep{chan+26}.

	\paragraph*{\bf Photomeson pair injection.}
	For larger magnetizations, photomeson interactions ($p\gamma$) inject pairs via $p+\gamma\rightarrow\pi^0+X$ and subsequent Breit--Wheeler absorption. The threshold is $\gamma_{p\gamma}\simeq 7.5\times 10^{4}(\varepsilon_X/1\,{\rm keV})^{-1}$ \citep{stecker68}, with fraction $f_{p\gamma}=n_{p,>\gamma_{p\gamma}}/n_{p,{\rm GeV}}\simeq 10^{-5}\gamma_{\min}(\varepsilon_X/1\,{\rm keV})$ of protons contributing. Letting above-threshold protons lose energy on photomeson timescales $t_{p\gamma}$ with $f_{\pi^0}\sim 1/3$ entering the $\pi^0$ channel, the pair loading relative to the GeV proton pool is:
	\begin{equation}\label{eq:npairs_over_npGeV_pg}
		\frac{n_\pm}{n_{p,{\rm GeV}}}
		\simeq
		\,f_{p\gamma}\,f_{\pi^0}\,
		\frac{E_p}{2m_ec^2}\,
		\min\!\left(\frac{t_{\rm esc}}{t_{p\gamma}},\,1\right)
		\sim
		3\times10^{2}\,
		\,\gamma_{\min}
	\end{equation}
	where the second equality is evaluated for $\varepsilon_X=1\,{\rm keV}$ and $E_{p,\rm th}/(2m_ec^2)\simeq 7\times10^{7}$. The dependence on $\ell_X$ is encompassed in the $t_{\rm esc}/t_{p\gamma} \simeq 1$ ratio \citep{mbarek26a}. This assumes that $\gamma$-rays cascade to MeV levels \citep[e.g.,][]{murase+20b}.
	
	\paragraph*{\bf{Synthesis}}
	Increasing $\sigma$ tends to produce harder nonthermal proton spectra \citep[e.g.,][]{zhdankin+17,zhdankin+18,comisso+18}, implying a larger population reaching the $p\gamma$ regime and higher pair loading $n_{\pm}/n_p$. Additionally, the pair magnetization $\sigma_e=B^2/(4\pi\bar{n}_e m_e c^2)$ decreases as the pair density (and hence $\tau_T$) rises at fixed heating. As pair creation proceeds, the pair density and $\tau_T$ rise while $\sigma_e$ falls, so the magnetization at coronal ``birth'' can exceed its pair-loaded equilibrium. Since $\ell_X$ depends on $L_X$ and $r_c$ rather than lepton density, this self-regulation acts on $\sigma_e$ and $T_e$, not on $\ell_X$ directly. Classical pair balance and observed compactness--temperature distributions make pair domination plausible for compact AGN coronae, but not universal \citep{svensson84,fabian+15,fabian+17}. It is worth emphasizing that the hadronic contribution above depends on the proton spectrum properties.

	\subsection{Flux-shedding power and coronal outflows}
	
	In the thin, wedge saturation layer where $f_r > 0$ just outside $r_c$, the ejected power is estimated as $\dot E_{\rm ej}\simeq 4\pi r_c^2 v_A f_r^{\max}\Delta r_{\rm eff}$. With peak radial force density $f_r^{\max}\simeq\kappa B^2/(8\pi r_c)$ ($\kappa=\mathcal{O}(1)$) and $\Delta r_{\rm eff}=\delta_{\rm s}r_c$:
	\begin{equation}\label{eq:ej-pwr}
		\dot E_{\rm ej} \simeq \frac{\kappa \delta_{\rm s}}{2} B^2 r_c^{2} c \sqrt{\frac{\sigma}{1+\sigma}},
	\end{equation}
	numerically:
	\begin{equation}\label{eq:ej-pwr-est}
		\begin{split}
			\dot E_{\rm ej} \simeq 3\times 10^{41}
			\left(\frac{\kappa}{1}\right)
			\left(\frac{\delta_{\rm s}}{0.1}\right)
			\left(\frac{B}{10^{3}\,\mathrm{G}}\right)^{2} \\
			\times \left(\frac{M}{10^{7}\,M_\odot}\right)^{2} 
			\left[\frac{\sigma}{1+\sigma}\right]^{1/2}
			\ \mathrm{erg\ s^{-1}}
		\end{split}
	\end{equation}
	where $\Delta r_{\rm eff}\lesssim\mathcal{O}(0.1\text{--}0.3)\,r_c$ for $\kappa\sim 1$ from the energy-budget bound. A fraction of the impulsive shedding can be converted into an extended coronal outflow magnetically connected to the compact X-ray corona, with continued dissipation at larger heights shaping longer-wavelength emission \citep{hankla+26,mbarek26a}.

	\subsection{Jet--corona power partition in flux-saturated states}
	
	\subsubsection{Why many radiative-mode systems remain weakly jetted}\label{sec:radio_quiet}
	
	The corona-to-jet partition follows from comparing $\lx$ to the BZ jet power \citep{blandford+77,tchekhovskoy+10}
	\begin{equation}
		P_{\rm BZ}
		=\frac{\kappa_{\rm BZ}}{4\pi}\phi_H^2\omega_H^2 f(\omega_H)
		\dot M_{\rm hor}c^2 ,
	\end{equation}
	where $\phi_H$ is the horizon-threading normalized flux,
	$\dot M_{\rm hor}$ is the mass flux reaching the horizon, $\omega_H = \Omega_H r_g/c$ is the dimensionless horizon frequency, and $f(\omega_H)$ is the high-spin correction to the BZ power. We distinguish $\phi_H$ from the coronal saturation normalization $\phi_c$ ($\phi_c=\phi$ denotes the coronal saturation flux defined in Eq.~\eqref{eq:phi_simplified}) entering Eq.~\eqref{eq:LX_Mc}. Combining with Eq.~\eqref{eq:LX_Mc} gives:
	\begin{align}
    \begin{split}
        \frac{P_{\rm BZ}}{L_X}
		\simeq
		3\,\mathcal{G}
		\left(\frac{\kappa_{\rm BZ}}{0.05}\right)
		\left(\frac{\omega_H}{0.5}\right)^2
		f(\omega_H)\\
		\times \left(\frac{r_c}{10r_g}\right)^2  \left(\frac{\phi_H}{\phi_c}\right)^2
		\left(\frac{\dot M_{\rm hor}}{\dot M_c}\right)
    \end{split}
	\end{align}
	where
	\begin{equation}
		\mathcal{G}=
		\kappa_{\rm x}
		\left(\frac{\mathcal{A}_X}{0.25}\right)^{-1}
		\left[\frac{(1+q_c^2)}{10}\right]^{-1},
		\quad
		\mathcal{A}_X = \cos\theta_f-\cos\theta_X ,
	\end{equation}
	collects the coronal factors, with $\mathcal{A}_X$ the effective angular covering factor of the X-ray corona. For $\theta_f\simeq10^\circ$--$30^\circ$ and an outer boundary $\theta_X\simeq45^\circ$, one has $\mathcal{A}_X\simeq0.16$--$0.28$, so the normalization $\mathcal{A}_X\simeq0.25$ is natural. From the definition of $\phi$, we can rewrite:
	\begin{equation}
		\left(\frac{\phi_H}{\phi_c}\right)^2
		\left(\frac{\dot M_{\rm hor}}{\dot M_c}\right)
		= \left(\frac{\Phi_H}{\Phi_c}\right)^2
	\end{equation}
	The partition can then be rewritten as: 
	\begin{equation}
		\frac{P_{\rm BZ}}{L_X}
		\simeq
		3\,\mathcal{G}
		\left(\frac{\kappa_{\rm BZ}}{0.05}\right)
		\left(\frac{\omega_H}{0.5}\right)^2
		f(\omega_H)
		\left(\frac{r_c}{10r_g}\right)^2
		\left(\frac{\Phi_H}{\Phi_c}\right)^2
		\label{eq:PBZ_LX_fH}
	\end{equation}
	
	We conclude that radio-quiet Seyferts and radiative-mode quasars correspond to modest spin, compact saturation, and incomplete horizon opening of the saturated flux.

	This weak-spin, weakly jetted interpretation is consistent with thin-disk GRMHD results \citep{avara+16}. For $r_c\simeq5$--$10\,r_g$ and $\omega_H\simeq0.2$--$0.4$, Eq.~\eqref{eq:PBZ_LX_fH} gives $P_{\rm BZ}/L_X\sim10^{-4}$--$0.4$ for $\Phi_H/\Phi_c\sim0.03$--$0.5$, within order-unity geometric factors. The upper end of this range, reached only for the largest $\Phi_H/\Phi_c$ and $r_c$, already approaches the radio-intermediate regime, while the bulk of the parameter space gives $P_{\rm BZ}/L_X\ll1$. Such systems can sustain a strong X-ray corona while launching a weak BZ jet.

	\subsubsection{Crossing into the radio-loud branch: spin, flux opening, and environment}
	
	Equation~\eqref{eq:PBZ_LX_fH} shows that a genuinely radio-loud radiative-mode source requires a coupled increase in spin, saturation radius, and open horizon-flux fraction. The dimensionless horizon frequency is bounded by $\omega_H=a_\ast/[2(1+\sqrt{1-a_\ast^2})]\le0.5$, with $f(\omega_H)$ of order unity and $\lesssim1$ toward high spin (e.g.\ $f\simeq0.77$ at $\omega_H=0.5$), so the spin factor is near-saturated and the remaining leverage comes from $r_c$ and $\Phi_H/\Phi_c$. At the lower edge of the relevant parameter box ($\omega_H\simeq0.5$, $r_c\simeq20\,r_g$, $\Phi_H/\Phi_c\simeq0.5$), Eq.~\eqref{eq:PBZ_LX_fH} gives only $P_{\rm BZ}/L_X\simeq2$. Reaching $P_{\rm BZ}/L_X\gtrsim10$ requires the upper end of every factor together (near-extremal spin $\omega_H\to0.5$, $r_c\sim25$--$30\,r_g$, and $\Phi_H/\Phi_c\sim0.7$--$1$) together with favorable geometric factors $\mathcal{G}\gtrsim1$. Such values are consistent with powerful radio-loud quasars and blazars whose jet powers are comparable to or larger than $\lx$.

	Spin and circumnuclear magnetization bias this transition further. High spin raises $\omega_H^2f(\omega_H)$ and may also push the saturation radius outward if ergospheric coupling supplies additional twist. The flux supply must also be coherent enough to give $\Phi_H/\Phi_c\sim1$. Very weakly magnetized inflow fails to build a strong bundle, whereas heavily mass-loaded starburst winds can build a compact X-ray corona but also keep the surrounding medium dense enough to frustrate weak jets. The radio-loud branch is therefore favored by high spin, coherent flux delivery, and a relatively low-density escape channel, as expected in massive elliptical or post-merger hosts.
	
	A second, environmental filter acts after launch.  
	The jet momentum flux/ram pressure $P_j$ in $\rm dyne\,cm^{-2}$:
	\begin{equation}
		P_j\sim \frac{L_j}{\pi\theta_j^2R_j^2c}
		\simeq 10^{-7}
		\left(\frac{L_j}{10^{43}\,{\rm erg\,s^{-1}}}\right)
		\left(\frac{\theta_j}{0.1}\right)^{-2}
		\left(\frac{R_j}{100\,{\rm pc}}\right)^{-2}
	\end{equation}
	where $\theta_j$ is the jet half-opening angle, $L_j$ the jet luminosity, and $R_j$ the distance from the black hole. A jet becomes an extended radio source only if its momentum flux exceeds the circumnuclear wind pressure. The same winds that help supply coherent magnetic flux to the inner flow can also frustrate weak jets on $R_{\rm w}\sim10$--$10^2\,{\rm pc}$ scales. For M82-like conditions, the wind ram pressure is
	$P_{\rm w}\sim10^{-8}$--$10^{-7}\,{\rm dyne\,cm^{-2}}$
	(Eq.~\eqref{eq:Pw}), comparable to the momentum flux of a weak Seyfert-like jet. Equating this with the wind pressure gives:
	\begin{equation}\label{eq:Lj_cret}
		L_{j,\rm crit}\sim5\times10^{42}
		\left(\frac{\theta_j}{0.1}\right)^2
		\left(\frac{\mathcal{M}{\rm SFR}}{10M_\odot{\rm yr^{-1}}}\right)
		\left(\frac{v_\infty}{10^3{\rm km\,s^{-1}}}\right)
		{\rm erg\,s^{-1}}
	\end{equation}
	For a Seyfert with $L_X\sim10^{43}$--$10^{44}{\rm erg\,s^{-1}}$, a jet with $P_{\rm BZ}/L_X\lesssim0.1$ can be disrupted or mass-loaded by a compact nuclear wind. By contrast, a radio-loud quasar with $P_{\rm BZ}/L_X\sim10$ launches $L_j\sim10^{44}$--$10^{45}{\rm erg\,s^{-1}}$ for the same X-ray luminosity scale and easily exceeds this propagation threshold. Thus radio quietness can arise either because the engine produces only a weak BZ jet, or because a modest jet is filtered by a dense circumnuclear environment: relativistic hydro/MHD simulations likewise find strong shocks, mixing, and deflection when weak jets encounter dense winds or clumpy circumnuclear gas \citep{bosch-ramon15,perucho+17,wagner+12,mukherjee+18a,mukherjee+18b,bicknell+18}. 
	
	\subsubsection{Synthesis}
	
	The picture above suggests that radio loudness in radiatively efficient AGN reflects a coupled topological transition rather than a single control parameter. Three ingredients are central: (i) the saturation radius $r_c$, (ii) the opening angle of the magnetically dominated wedge, and (iii) importantly, the fraction of coherent flux that reaches the horizon. 
	
	An increase in $r_c$, which could be associated with a higher spin (\S\ref{sec:rc_def}), has an important radiative consequence: it need not make the corona brighter. Hard X-ray efficiency then depends not only on magnetic heating, but also on compactness, optical depth, and Comptonization, with $\ell=L\sigma_T/(Rm_ec^3)$, $y\sim4\theta_e\max(\tau_T,\tau_T^2)$, and $\lambda_{\rm cool}/R\simeq3\pi/(\ell\gamma)$. Compact regions with $\ell\gg1$ and $y\sim1$ radiate electron heating efficiently as Comptonized X-rays. If the emitting region expands, $\ell\propto L/R$ and often $\tau_T$ decrease, photons scatter less, and the corona can move away from the efficient thermal-Compton regime. Radio loudness can therefore correspond to a larger coherent flux channel to the hole, but a more diluted or outward-shifted X-ray emitting region.

	The circumnuclear flux supply sets the environmental filter for this topology. At fixed disk thermodynamics, Eq.~\eqref{eq:betaLS_from_betaw} gives $\beta_{\rm g}(r)\propto\beta_w$, hence $\Pi_{\rm ord}=\Xi\beta_{\rm g}\propto\beta_w$. Strongly magnetized winds lower $\Pi_{\rm ord}$, widen the wedge through Eq.~\eqref{eq:ff_wedge_condition}, and favor a compact, corona-dominated topology in which much of the coherent flux is trapped outside the horizon. More moderately magnetized winds keep $\Pi_{\rm ord}(r_c)$ closer to unity, narrow the wedge, and allow more structured disk flux to connect to the hole. The dependence is therefore non-monotonic: the flux supply must build the inner bundle without over-truncating the disk or trapping most of the coherent flux in the corona.
	
	Radio loudness is then favored by high spin, coherent flux delivery, moderate circumnuclear magnetization, and a low-density escape channel. High-spin, high-$P_{\rm BZ}$ states are likely associated with coherent or merger-driven accretion, whereas chaotic secular feeding tends to produce more modest spins, $a_\ast\sim0.1$--$0.3$ \citep{volonteri+07,king+08,volonteri+13}.
	At fixed $\lambda_{\rm E}$, it should correlate statistically with spin, host gas content, and merger history, with compact nuclear starbursts under-represented and post-starburst or merger signatures enhanced, broadly pointing to massive elliptical or post-merger hosts rather than gas-rich compact starbursts \citep{sikora+07,best+05,chiaberge+15}. This is a statistical statement, not a strict dichotomy: radio-loud, jetted systems also occur in gas-rich, star-forming hosts, including radio-loud and $\gamma$-ray narrow-line Seyfert~1s in disk-dominated galaxies and young compact CSS/GPS sources embedded in dense circumnuclear gas \citep{odea+21}. 
	
	X-ray reflection provides an independent diagnostic of the same topology. Jetted AGN with $\lambda_{\rm E}\gtrsim0.01$ may have larger effective X-ray emitting regions \citep{dogruel+20}, which dilute disk illumination, weaken reflection, and increase reverberation lags \citep{fabian+00}. Weaker Fe K/reflection features are often reported in radio-loud AGN \citep[e.g.,][]{sambruna+99}, although not universally \citep{zhu+21}. Consistently, broad-line radio galaxies and radio-loud quasars in elliptical hosts show Comptonizing coronae with measured high-energy cutoffs and X-ray variability \citep{ballantyne+14,lohfink+15,kang+20}, while radio-loud coronae appear systematically cooler than their radio-quiet counterparts \citep{kang+22}.


    \subsection{Little Blue and Red Dots as candidate precursors to radiative-mode AGN}

    The model predicts a population of precursors with weak X-ray emission, in which disk reservoirs around black holes with $M_\bullet \geq 10^6M_{\odot}$ have yet to accumulate enough coherent magnetic flux from circumnuclear stellar winds to sustain a luminous corona. Little Blue Dots (LBDs) and Little Red Dots (LRDs) could belong to this population. These compact sources show broad Balmer emission consistent with accreting SMBHs, with LRDs distinguished by their red optical continua and characteristic V-shaped SEDs, while LBDs have bluer continua \citep{Kocevski+25,rinaldi+25,brazzini+26}. Other work instead attributes their properties to super-Eddington accretion \citep{kaaz+26,king26}. The reservoir phase proposed here is a complementary rather than exclusive interpretation. Most LRDs remain undetected in \emph{Chandra} at levels well below expectations for AGN with broad emission lines \citep{ananna+24,akins+25}, and strong X-ray weakness is also observed in representative LBDs \citep{brazzini+26}. In this model, either class could host a reservoir awaiting sufficient coherent inner flux to sustain a luminous corona.

    X-ray weakness alone is not decisive. Heavy obscuration can hide a corona \citep{ricci+18}, while declining accretion, reduced cooling by seed photons, or changes in obscuration can reveal or reshape one. The clearest tests would distinguish how star formation feeds the inner flow. In this model, cold gas accretion preferentially builds the disk reservoirs associated with LRDs and LBDs, whereas magnetized winds powered by stellar activity and supernovae supply the coherent flux needed to sustain luminous coronae in Seyferts and quasars. The relevant distinction is the inward delivery of coherent magnetic flux rather than field strength alone. In weakly ionized gas, ambipolar diffusion lets the field drift relative to the neutral inflow and can impede coherent flux transport \citep{bai+17}, lowering $f_\Phi$. Hot, well-ionized winds couple more effectively. These populations should therefore trace star formation differently, according to how much cold gas and coherent magnetic flux reach the inner regions, rather than the total star formation rate alone. At comparable gas supply, stronger inward delivery of coherent wind flux should favor coronal emergence, with intermittent hard X-ray emission as flux accumulates and changes in coronal size accompanying variability \citep{wilkins+12}.

	\balance
	\section{Conclusions}
	
	We have developed a physically motivated pathway for the onset of radiative-mode AGN in which circumnuclear star formation supplies coherent magnetic flux to a compact disk reservoir around the SMBH. If a net poloidal component survives inward advection, it can become dynamically important near the black hole, compress the disk atmosphere, and open a magnetically dominated wedge. Within this wedge, the large-scale field self-regulates toward a critical twist at which the net radial transport changes sign, defining a magnetic saturation radius $r_c\sim10\,r_g$. The resulting flux-regulated inner bundle provides a natural site for compact, time-dependent coronal heating, while the same topology controls how electromagnetic power is partitioned between local X-ray dissipation, coronal outflows, and relativistic jets. 

	Previous work has linked net poloidal flux to magnetic support, disk truncation, and corona formation \citep{miller+00,salvesen+16,liska+22,scepi+24,bambic+24a,dhang+25,zhang+25}. The distinct step here is to connect those ingredients to a wind-fed flux supply, a twist-defined $r_c$, and observable coronal scalings. Our main conclusions are:
	
	\begin{itemize}
		\item \textit{Circumnuclear winds provide a plausible flux-supply channel.}
		Stellar- and supernova-driven winds from nuclear star formation can deliver partially ordered magnetic flux into the SMBH fueling region. Whether this flux affects the inner engine depends on the survival of a coherent net poloidal component during capture and inward transport.
		
		\item \textit{Ordered fields trigger a transport-regime transition.}
		As the ordered component strengthens and the plasma beta drops, mean Maxwell stresses can overtake MRI turbulence, shifting angular-momentum transport from turbulence-dominated to large-scale-field-dominated and enabling a magnetically dominated wedge above the disk.
		
		\item \textit{Critical twist sets a magnetic saturation radius.}
		In a quasi-force-free description of the wedge, an effective radial-transport diagnostic changes sign when the twist reaches a topology-dependent value $q_c(p,g)$. This defines a magnetic saturation radius $r_c$ set mainly by the radial slopes of the poloidal and toroidal field components, with natural values near the observed coronal scale, $r_c\sim10\,r_g$.
		
		\item \textit{Flux regulation fixes the inner magnetic structure.}
		Additional flux advected toward $r_c$ enhances the outward electromagnetic response and is preferentially shed. The time-averaged result is an approximately constant enclosed radial flux for $r\le r_c$, implying $B_r\propto r^{-2}$ inside the saturated bundle and a normalized flux comparable to the MAD range found in GRMHD simulations. Flux saturation can persist without strong disk squeezing. Adopting $\Pi_c\simeq1$ instead gives $\phi\simeq10$--$20$ for the fiducial parameters.
		
		\item \textit{Flux shedding supplies a natural coronal heating and variability channel.}
		The balance between flux advection and flux shedding makes the saturation boundary an impulsive forcing site. Reconnection and turbulence driven there can provide intermittent heating, with the coronal luminosity and spectrum controlled by $r_c$, compactness, optical depth, and magnetization.
		
		\item \textit{The corona--jet partition is topological.}
		Field geometry, spin coupling, wedge opening, and the fraction of coherent flux reaching the horizon determine whether the incoming electromagnetic power is dissipated locally in the corona or carried away by a BZ jet. Radio quietness can therefore reflect either a weak intrinsic jet or environmental filtering by dense circumnuclear gas.
		
		\item \textit{Little Blue and Little Red Dots are candidate precursor systems.}
		The model admits an X-ray-weak reservoir phase with an accreting SMBH and star-forming nucleus before a luminous corona emerges. Little Red Dots are candidate precursors, and rare X-ray-bright outliers may be transitional.
		
	\end{itemize}
	
	The broader implication is that AGN X-ray emergence may be controlled not only by the mass supply rate, but also by the topology, coherence, and survival of magnetic flux delivered from the host environment. This provides a link between circumnuclear star formation, coronal variability, radio loudness, and possible high-redshift precursor populations. Crucially, the wind-fed channel and the saturation physics are logically distinct: circumnuclear winds act as the switch that sets whether and when coherent flux accumulates, whereas the topology of the resulting quasi-force-free wedge sets the state and depends on a generic supply of ordered flux. The saturation radius, normalized flux, and coronal scalings derived here should therefore carry over to any channel that delivers ordered poloidal flux to the inner flow, including an internal disk dynamo. Although the global flux-feeding channel is for SMBHs, the saturation physics of magnetized bundles may also be relevant to X-ray binaries.
	
	\section*{Acknowledgements}
	I thank the anonymous referee for constructive comments, Elias Most for reviewing the manuscript, and Nick Kaaz, Eliot Quataert, Alexander Philippov, Anatoly Spitkovsky, and Chris Reynolds for useful discussions and comments. 
    R.M. is supported by a Lyman Spitzer Fellowship at Princeton University.

    \section*{Data Availability} No new data were generated or analysed in support of this research.
	
	\balance
	\bibliography{Total,Total_cont}
	\bibliographystyle{mnras}
	\balance

	\appendix
	\section{Radial Flux Bundle}
	\label{app:flux-scaling}
	
	Working in spherical coordinates $(r,\theta,\phi)$ with axisymmetry, take the poloidal field to be purely radial, $\mathbf{B}_p = B_r(r,\theta)\hat{r}$. In ideal MHD, flux through any comoving fluid loop is conserved, so for a bundle of fluid elements advected inward, $\Phi_{\rm bundle}=\text{const}$. At radius $r$, the bundle spans $\theta_1(r)\leq\theta\leq\theta_2(r)$, with solid angle $\Delta\Omega(r) = 2\pi[\cos\theta_1(r) - \cos\theta_2(r)]$. If lateral mixing is weak, each fluid element falls in along an approximately fixed $\theta$, i.e. $|v_\theta/v_r|\ll 1$ implies $d\theta_i/dr\approx 0$ and hence $\Delta\Omega\approx\text{const}$. The flux through a spherical cap at $r$ spanning the bundle is then
	\begin{equation}
		\Phi_{\rm bundle} = \int B_r\, r^2\sin\theta\, d\theta\, d\phi \approx B_r\, r^2\,\Delta\Omega,
	\end{equation}
	where we have assumed $B_r$ is roughly uniform across the (narrow) bundle. Combining with $\Phi_{\rm bundle}=$ const and $\Delta\Omega\approx$ const gives $B_p = B_r \propto r^{-2}$. The result rests on three assumptions: (i) $\Phi_{\rm bundle}$ conserved, (ii) weak lateral mixing ($\Delta\Omega$ conserved), and (iii) predominantly radial field ($B_p\approx B_r$).
	
	\section{Column fraction above the wedge--disk interface}
	\label{app:column_fraction}
	
	The criterion for magnetic domination of a substantial column fraction (Eq.~\eqref{eq:squeezing_onset}) rests on a geometric statement: Squeezing acts mainly on the gas column above the wedge--disk interface. Here we derive the explicit form of that column fraction.
	
	\paragraph*{Isothermal vertical structure.}
	For a thin, vertically isothermal disk with $P=\rho c_s^2$ and
	vertical gravity $g_z=-\Omega_K^2 z$, hydrostatic equilibrium gives
	\begin{equation}
		\rho(z)=\rho_0\,e^{-z^2/2H^2},
		\qquad H= c_s/\Omega_K,
	\end{equation}
	and total surface density
	$\Sigma_{\rm total}=\rho_0\sqrt{2\pi}\,H$. The column above a height
	$z_{\rm int}=n_{\rm int}H$ is
	\begin{equation}
		\Sigma(>z_{\rm int})
		=\int_{z_{\rm int}}^{\infty}\rho(z)\,dz
		=\rho_0 H\sqrt{\pi/2}\;
		\mathrm{erfc}\!\left(\frac{n_{\rm int}}{\sqrt{2}}\right),
	\end{equation}
	where the change of variable $u=z/(\sqrt{2}H)$ and the definition $\mathrm{erfc}(x)=(2/\sqrt{\pi})\int_x^\infty e^{-u^2}du$ have been used. The exact fractional column above the interface is therefore
	\begin{equation}\label{eq:column_exact}
		\frac{\Sigma(>z_{\rm int})}{\Sigma_{\rm total}}
		=\tfrac{1}{2}\,\mathrm{erfc}\!\left(\frac{n_{\rm int}}{\sqrt{2}}\right).
	\end{equation}
	
	\paragraph*{Gaussian-tail asymptotic.}
	For $n_{\rm int}\gtrsim 1$, the asymptotic expansion
	$\mathrm{erfc}(x)\simeq e^{-x^2}/(x\sqrt{\pi})$ gives
	\begin{equation}
		\frac{\Sigma(>z_{\rm int})}{\Sigma_{\rm total}}
		\simeq\frac{1}{\sqrt{2\pi}\,n_{\rm int}}\,e^{-n_{\rm int}^2/2}
		\sim e^{-n_{\rm int}^2/2}.
		\label{eq:column_asymptotic}
	\end{equation}
	The Gaussian factor dominates. The $1/n_{\rm int}$ prefactor is kept implicit in the approximation.
	
	\paragraph*{Relating to $\Pi_{\rm ord}$.}
	The interface height is not free: it is fixed by the pressure
	balance $P_g(z_{\rm int})=P_{\rm B}$. For an isothermal atmosphere,
	$P_{\rm sup}(z)=P_{\rm mid}e^{-z^2/2H^2}$, with $\Xi$ slowly varying as in \S\ref{sec:ang_extent}, so
	\begin{equation}
		\Pi_{\rm ord}=\frac{P_{\rm mid}}{P_{\rm B}}
		=e^{+n_{\rm int}^2/2}
		\quad\Longleftrightarrow\quad
		n_{\rm int}=\sqrt{2\ln\Pi_{\rm ord}},
	\end{equation}
	which recovers Eq.~\eqref{eq:ff_wedge_condition} evaluated at the interface. 

    Using the exact form in Eq.~\eqref{eq:column_exact} gives:
\begin{equation}\label{eq:column_beta}
    \frac{\Sigma(>z_{\rm int})}{\Sigma_{\rm total}}
    \simeq
    \frac{1}
    {2\sqrt{\pi\ln\Pi_{\rm ord}}\,\Pi_{\rm ord}}
    \qquad (\Pi_{\rm ord}\gg1).
\end{equation}
The fractions of the total column above the interface on one side are $0.12\%$, $1.6\%$, and $50\%$ for $\Pi_{\rm ord}=100$, $10$, and $1$, respectively. At $\Pi_{\rm ord}=1$ the interface reaches the midplane, where each side contains half the total column. The layer therefore changes from a thin atmospheric skin to a
substantial part of the disk body as $\Pi_{\rm ord}$
approaches unity.
    
	
	
	\section{Horizon regularity and the inner twist}
\label{app:qH}

The Znajek condition ensures regularity of a stationary, axisymmetric electromagnetic field at the future horizon \citep{znajek77,penna15}. In the ZAMO orthonormal frame, its near-horizon form gives
\begin{equation}
\frac{|B_{\hat\phi}|}{|B_{\hat r}|}
\simeq
\frac{|\Omega_H-\Omega_F|\,\varpi_H}{\alpha_k c},
\qquad r\rightarrow r_H^+,
\end{equation}
The Boyer--Lindquist lapse $\alpha_k\propto\sqrt{\Delta}$
vanishes at the horizon, where $\Delta=r^2-2r_gr+a_\ast^2r_g^2=0$. At an exterior footpoint $r_0$, we assume comparable differential-winding and Alfv\'enic relaxation times, so the generated toroidal field is comparable to the radial field, motivating $q_0\simeq1$.
	
	\section{Intermediate value theorem and the existence of a magnetized bundle boundary}
	\label{app:ivt}
	
	The radius $r_c$ is defined by $f_r(r_c)=0$, equivalently the radius where $q(r_c)=q_c$ with $\mathcal{F}(q_c)=0$. For smooth fields, $q(r)$ is continuous on any exterior interval $[r_1,r_2]\subset(r_H,\infty)$, hence $f_r(r)$ is also continuous.
	
	Defining $y_r(r)= q(r)-q_c$, an exterior $r_c$ exists if and only if there exist $r_1,r_2>r_H$ with $[q(r_1)-q_c][q(r_2)-q_c]<0$. By the intermediate value theorem, $y_r(r)$ then vanishes somewhere between $r_1$ and $r_2$, giving $r_c$ with $f_r(r_c)=0$.

	For $g>p$, the adopted profile decreases outward from $q_0$ at $r_0$. If $q_0<q_c$, then $q(r)<q_c$ for $r\ge r_0$, so no saturation radius occurs in this interval. This conclusion follows from the exterior normalization and does not constrain the region $r_H<r<r_0$. Conversely, if $q$ exceeds $q_c$ somewhere outside the horizon, the bracketing condition is met and at least one $r_c$ is guaranteed. For $g>\tfrac12$, $\mathcal{F}(q)$ has a single positive root, so a monotonic inward crossing of $q_c$ produces a unique $r_c$, and once $q>q_c$ one has $f_r>0$ at smaller radii.

	\section{Derivation of the polar EM force}\label{app:ftheta_derivation}
	
	In spherical coordinates $(r,\theta,\phi)$ with axisymmetry, the
	Maxwell stress tensor for a field with
	$B_\theta\ll(B_r^2+B_\phi^2)^{1/2}$ has diagonal components
	\begin{align}
		T_{rr}   = \frac{B_r^2-B_\phi^2}{8\pi}, \quad
		T_{\theta\theta} = -\frac{B_r^2+B_\phi^2}{8\pi}, \quad
		T_{\phi\phi}     = \frac{B_\phi^2-B_r^2}{8\pi},
	\end{align}
	and off-diagonal $T_{r\phi}=B_rB_\phi/(4\pi)$. The
	$\theta$-component of the divergence of the Maxwell stress tensor in
	spherical coordinates is
	\begin{equation}
		f_\theta = \frac{1}{r}\partial_\theta T_{\theta\theta}
		+ \frac{\cot\theta}{r}(T_{\theta\theta}-T_{\phi\phi})
		+ \partial_r T_{r\theta} + \frac{3}{r}T_{r\theta}.
	\end{equation}
	The last two terms vanish for $B_\theta\simeq 0$. Substituting,
	$T_{\theta\theta}-T_{\phi\phi}=-B_\phi^2/(4\pi)$ and
	$\partial_\theta T_{\theta\theta}=-\partial_\theta(B_r^2+B_\phi^2)/(8\pi)$,
	giving the full polar force density
	\begin{equation}\label{eq:ftheta_full_app}
		f_\theta
		= -\frac{1}{8\pi r}\partial_\theta(B_r^2+B_\phi^2)
		-\frac{\cot\theta}{4\pi r}\,B_\phi^2.
	\end{equation}
	The first term is the magnetic pressure gradient, nonzero only where $B^2$ varies with $\theta$. The second is the toroidal hoop stress, present wherever $B_\phi\neq 0$.
	
	\paragraph*{Angular regimes.}
	The wedge condition (Eq.~\eqref{eq:ff_wedge_condition}),
	$|\cos\theta|>\epsilon\sqrt{2\ln\Pi_{\rm ord}}$, defines the polar angle $\theta_d\simeq\pi/2-n_{\rm int}\epsilon$ (with $n_{\rm int}=\sqrt{2\ln\Pi_{\rm ord}}$) marking $P_{\rm sup}=P_B$, where $P_{\rm sup}=P_{\rm g}+P_{\rm rad}$. Eq.~\eqref{eq:ftheta_full_app} then organizes into the two regimes below (the funnel wall is treated in \S\ref{sec:polar}).
	
	\emph{Wedge body ($\theta<\theta_d$, away from the funnel wall).} In the wedge interior, we assume that magnetic pressure varies weakly with $\theta$, so $\partial_\theta B^2\simeq0$ and the polar force is dominated by hoop stress:
	\begin{equation}
		f_\theta\simeq -\frac{B_\phi^2\cot\theta}{4\pi r}
		= -\frac{q^2 B_r^2\cot\theta}{4\pi r},
		\qquad \theta<\theta_d.
	\end{equation}
	This is negative in the upper hemisphere by the sign of $\cot\theta$ alone, independent of the field profile: the wedge body collimates the flow poleward.
	
	\emph{Disk--wedge interface.}
    The ordered radial field declines toward the midplane across $\Delta\theta\sim\epsilon$. As assumed in \S\ref{sec:polar}, differential winding of MRI-driven radial fluctuations approximately compensates for the declining ordered toroidal pressure, leaving a weak gradient in the total toroidal magnetic pressure. The radial field therefore supplies the dominant squeezing contribution:
\begin{equation}
|f_\theta^{\rm sq}|
\sim \frac{|\partial_\theta B_r^2|}{8\pi r}
\sim \frac{B_r^2}{8\pi r\epsilon},
\end{equation}
directed toward the midplane from both sides. Relative to the toroidal hoop stress,
\begin{equation}
\frac{|f_\theta^{\rm sq}|}{|f_\theta^{\rm wb}|}
\sim \frac{1}{2q^2\epsilon|\cot\theta|}.
\end{equation}
Using $\theta_d$ as the reference interface angle,
$\cot\theta_d\simeq n_{\rm int}\epsilon$, this ratio becomes
$\sim[2q^2n_{\rm int}\epsilon^2]^{-1}$. When this ratio exceeds unity, the net polar force changes from collimating in the wedge body to squeezing near the disk
(Figure~\ref{fig:ftheta_sign}).

	\section{Derivation of the saturation flux $\phi$}\label{app:sat_flux_der}
	
	We adopt the standard horizon-flux normalization
	\citep{tchekhovskoy+11,narayan+22}, defining the magnetic flux through
	one hemisphere of a sphere of radius $r$ in the saturated bundle as
	\begin{equation}\label{eq:phi_de_cemi}
		\Phi_c = \frac{1}{2}\int_0^\pi\!\!\int_0^{2\pi}
		|B^r(r_c,\theta,\phi)|\,r_c^2\sin\theta\,d\theta\,d\phi
		\;\simeq\; 2\pi r_c^2\,\eta_r\,B(r_c),
	\end{equation}
	where $B(r_c)$ is the total field magnitude in the wedge,
	$\eta_r= (1+q_c^2)^{-1/2}$ extracts the radial component
	$B_r=\eta_r B$ from the toroidally wound field, and the factor of
	$1/2$ averages over the two hemispheres of opposite polarity. 
	
	Working with Eq.~\eqref{eq:confinement},
	\begin{align}
		\frac{B^2(r_c)}{8\pi}
		&\simeq \xi\,\rho(r_c)\,v_K^2(r_c),\\[3pt]
		\rho(r_c)
		&\simeq \frac{\dot M_c}{4\pi r_c^2 v_r(r_c)},\qquad
		v_r(r_c)=\zeta\,v_K(r_c),
	\end{align}
	with $v_K(r)=\sqrt{GM_\bullet/r}=c\sqrt{r_g/r}$. Combining gives
	\begin{equation}
		B^2(r_c)
		\simeq \frac{2\xi\,\dot M_c\,v_K^2(r_c)}{r_c^2\,v_r(r_c)}
		=\frac{2\xi}{\zeta}\,\frac{\dot M_c\,v_K(r_c)}{r_c^2}.
	\end{equation}
	Substituting into Eq.~\eqref{eq:phi_de_cemi} and normalizing by
	$(\dot M_c r_g^2 c)^{1/2}$ yields
	\begin{align}
		\phi
		&= \frac{\Phi_c}{(\dot M_c r_g^2 c)^{1/2}}
		\;\simeq\; 2\sqrt{2}\,\pi\,\eta_r\,
		\sqrt{\frac{\xi}{\zeta}}\,
		\frac{r_c}{r_g} \sqrt{\frac{v_K(r_c)}{c}}\nonumber\\[3pt]
		&\simeq\; 2\sqrt{2}\,\pi\,\eta_r\,
		\sqrt{\frac{\xi}{\zeta}}\,
		\left(\frac{r_c}{r_g}\right)^{3/4} ,
	\end{align}
	using $v_K/c=\sqrt{r_g/r_c}$. Inserting the topological scaling
	$r_c/r_g\simeq q_c^{1/(p-g)}$,
	\begin{equation}\label{eq:phi_full_app}
		\phi
		\simeq 2\sqrt{2}\,\pi\,\eta_r\,
		\sqrt{\frac{\xi}{\zeta}}\,
		q_c^{\frac{3}{4(p-g)}} ,
	\end{equation}
	which is plotted in Figure~\ref{fig:flux_sat} for different values of $p$ and $g$.

	\section{Electromagnetic versus gravitational force}
	\label{app:force_ratio_gr}
	
	\paragraph*{GR correction factor.}
	We normalize the electromagnetic estimates by the magnitude of the locally measured radial acceleration of a zero-angular-momentum observer (ZAMO), which remains at fixed $r$ and $\theta$ while rotating with the local frame dragging,
\begin{equation}
g_r^{\rm Kerr}
= |g_{\hat r}^{\rm ZAMO}|
=
\frac{c^2}{\sqrt{g_{rr}}}
\left|\partial_r\ln\alpha_k\right|,
\end{equation}
where $\alpha_k$ is the Kerr lapse and $g_{rr}$ is the
Boyer--Lindquist radial metric component. Thus,
\begin{equation}
\mathcal{R}_{\rm GR}(r,a_\ast)
=
\frac{c^2}{r\,g_r^{\rm Kerr}}
=
\frac{\sqrt{g_{rr}}}
     {r\,|\partial_r\ln\alpha_k|}.
\end{equation}
In Schwarzschild,
\begin{equation}
\mathcal{R}_{\rm GR}(r,0)
=
\frac{r}{r_g}\sqrt{1-\frac{2r_g}{r}},
\end{equation}
giving $\mathcal{R}_{\rm GR}=\sqrt{80}\simeq 9$
at $r=10r_g$. We retain the flat-space magnetic-force
estimates and use this acceleration as a reference
normalization. In the equatorial plane, with $G=M=c=1$ and $r$ measured in units of $r_g$,
	\begin{equation}
		\partial_r\ln\alpha_k
		=
		\frac{1}{2}
		\left(
		\frac{\Delta'}{\Delta}
		+
		\frac{2}{r}
		-
		\frac{A_{\rm k}'}{A_{\rm k}}
		\right),
	\end{equation}
	with
	\begin{equation}
		\alpha_k^2 =
		\frac{\Delta r^2}{A_{\rm k}},
		\quad
		\Delta=r^2-2r+a_\ast^2,
		\quad
		A_{\rm k}=r^4+a_\ast^2r^2+2a_\ast^2r ,
	\end{equation}
	and
	\begin{equation}
		\Delta'=2r-2,
		\qquad
		A_{\rm k}'=4r^3+2a_\ast^2r+2a_\ast^2
	\end{equation}
	The equatorial radial metric component is $g_{rr}=r^2/\Delta$. This expression is used to evaluate $\mathcal{R}_{\rm GR}$ for each spin $a_\ast$. Far from the hole, $\mathcal{R}_{\rm GR}\rightarrow r/r_g$, while near the ISCO and horizon the spin dependence becomes important.
	
	\paragraph*{The $f_r$ term.}
	The radial component is useful because its sign determines whether the electromagnetic force pushes the wedge inward or outward. For the power-law field model $B_r\propto r^{-p}$ and $B_\phi\propto r^{-g}$, the radial force
	can be written
	\begin{equation}
		f_r \simeq \frac{B_r^2}{4\pi r}\,\mathcal{F}(q),
		\qquad q= \frac{B_\phi}{B_r}.
	\end{equation}
	Using $B^2=B_r^2(1+q^2)$ and
	$\sigma= B^2/(4\pi\rho c^2)$ gives the signed ratio
	\begin{equation}
		\label{eq:ratio_radial_signed}
		\frac{f_r}{|\rho g_r|}
		\simeq
		\frac{\sigma\,\mathcal{R}_{\rm GR}}{1+q^2}\,\mathcal{F}(q),
	\end{equation}
	so radial EM stresses become dynamically important when
	\begin{equation}
		\label{eq:sigma_crit_radial}
		\sigma \gtrsim
		\frac{1+q^2}{\mathcal{R}_{\rm GR}(r,a_\ast)\,|\mathcal{F}(q)|}.
	\end{equation}
	For $q=2$, $p=2$, $g=1.5$, and $r=10r_g$ in Schwarzschild, $\mathcal{R}_{\rm GR}=\sqrt{80}$ and $\mathcal{F}(2)=-1$, giving $\sigma\gtrsim5/\sqrt{80}\simeq0.5$.
	
	\paragraph*{Wedge-body polar EM force.}
	The polar force relevant for the coronal body is not the thin-interface squeezing force. In the body of the magnetized wedge, the magnetic pressure is approximately uniform in angle, so the pressure-gradient term is subdominant.
	The remaining polar force is the toroidal hoop stress,
	\begin{equation}
		f_\theta^{\rm wb}
		\simeq
		-\frac{\cot\theta}{4\pi r}B_\phi^2 ,
	\end{equation}
	which is directed poleward. Writing
	$B_\phi^2=B^2q^2/(1+q^2)$ gives
	\begin{equation}
		\label{eq:ratio_wedge_body}
		\frac{|f_\theta^{\rm wb}|}{|\rho g_r|}
		\simeq
		\sigma\,\mathcal{R}_{\rm GR}(r,a_\ast)
		\frac{q^2}{1+q^2}\cot\theta .
	\end{equation}
	Electromagnetic stresses dominate the wedge-body polar dynamics when
	\begin{equation}
		\label{eq:sigma_crit_wedge_body}
		\sigma
		\gtrsim
		\frac{1+q^2}
		{\mathcal{R}_{\rm GR}(r,a_\ast)\,q^2\cot\theta}.
	\end{equation}
	For example, at $\theta=45^\circ$, $q=2$, and $r=10r_g$
in Schwarzschild, $\mathcal{R}_{\rm GR}=\sqrt{80}$,
giving $\sigma\gtrsim5/(4\sqrt{80})\simeq0.14$.
	


\section{Radiation-to-gas pressure ratio}
\label{app:prad_pgas}
We use the radiation-dominated, electron-scattering branch of
\citet[][Chapter~5, Eqs.~5.41, 5.47 and 5.58]{frank+02}, with its one-zone convention $\rho\simeq\Sigma/H$ and
$\nu=\alpha\Omega_KH^2$. Defining
$\mathcal J=1-\sqrt{r_{\rm in}/r}$ with $r_{\rm in}$ the disk inner edge, the midplane quantities satisfy
\begin{equation}
H=\frac{3\kappa_{\rm es}\dot M\mathcal J}{8\pi c},
\qquad
T^4=\frac{8c\Omega_K}{3a\alpha\kappa_{\rm es}},
\qquad
\rho=\frac{\dot M\mathcal J}{3\pi\alpha\Omega_KH^3},
\end{equation}
where $a$ is the radiation constant and $\kappa_{\rm es}=0.34$ the electron-scattering opacity. Hence
$P_{\rm rad}/P_{\rm g}=a\mu m_pT^3/(3\rho k_{\rm B})$ gives
\begin{equation}
\begin{aligned}
\frac{P_{\rm rad}}{P_{\rm g}}\simeq{}&8.4
\left(\frac{\alpha}{0.1}\right)^{1/4}
\left(\frac{M_\bullet}{10^7M_\odot}\right)^{1/4}
\left(\frac{\lambda_{\rm E}/\eta_{\rm rad}}{0.1}\right)^2\\
&\times\left(\frac{r}{10r_g}\right)^{-21/8}
\left(\frac{\mathcal J}{0.2254}\right)^2,
\end{aligned}
\end{equation}
for $\mu=0.62$, $\kappa_{\rm es}=0.34\,{\rm cm^2\,g^{-1}}$,
$\dot M_{\rm tot}=\lambda_{\rm E}L_{\rm Edd}/(\eta_{\rm rad} c^2)$, and
$L_{\rm Edd}=4\pi GM_\bullet m_pc/\sigma_T$.
At $r=10r_g$, $r_{\rm in}=6r_g$, $\alpha=\eta_{\rm rad}=0.1$, and
$\lambda_{\rm E}=0.01$, this gives $P_{\rm rad}/P_{\rm g}\simeq5$--15
for $M_\bullet=10^6$--$10^8M_\odot$, increasing as
$\lambda_{\rm E}^2$. These reference midplane estimates motivate retaining radiation support. The interface ratio $\Xi_c$ must be evaluated locally.

\section{Radiation-supported wedge base}
\label{app:rad_wedge}

As an illustrative alternative to the isothermal gas Gaussian, we consider a polytropic layer,
$P_{\rm sup} \propto \rho^\Gamma$ with $\Gamma=1+1/n_{\rm pol}$. Hydrostatic balance,
$dP_{\rm sup}/dz=-\rho\Omega_K^2z$, gives
\begin{equation}
    \frac{P_{\rm sup}(z)}{P_{\rm sup}(0)}
    =\left(1-\frac{z^2}{z_0^2}\right)^{n_{\rm pol}+1},
    \qquad
    z_0^2=\frac{2(n_{\rm pol}+1)P_{\rm sup}(0)}
    {\rho_0\Omega_K^2},
\end{equation}
for $|z|\leq z_0$, where $\rho_0=\rho(0)$.
Here $z_0$ is the zero-pressure surface. A radiation-dominated layer has $\Gamma=4/3$, hence $n_{\rm pol}=3$. 

With $B^2$ again weakly dependent on $\theta$,
$\Pi(r,\theta)=\Pi_{\rm ord}(r)\,[1-(r\cos\theta/z_0)^2]^{n_{\rm pol}+1}$ within the
layer. Magnetic domination, $\Pi(r,\theta)<1$, then holds for
\begin{equation}\label{eq:rad_wedge_condition}
    |\cos\theta|>
    \frac{z_0}{r}\left(1-\Pi_{\rm ord}^{-1/(n_{\rm pol}+1)}\right)^{1/2}
    \qquad(\Pi_{\rm ord}\geq1),
\end{equation}
while for $\Pi_{\rm ord}<1$ the bracket is negative and $\Pi<1$ throughout
the layer, down to the midplane.
For $n_{\rm pol}=3$, the interface heights are
$z_{\rm int}/z_0=0.40$, $0.66$, and $0.83$ at
$\Pi_{\rm ord}=2$, $10$, and $100$, respectively.
As $\Pi_{\rm ord}\to\infty$, the interface approaches
$z_0$, whereas the formal Gaussian height grows as
$H\sqrt{2\ln\Pi_{\rm ord}}$.
In either profile, decreasing $\Pi_{\rm ord}$ moves
the interface toward the midplane and widens the wedge.
For $\Pi_{\rm ord}<1$, magnetic pressure exceeds
the support pressure throughout the reference layer.

\section{Thin-disk thickness and cooling at $r_c$}
\label{app:thin_disk}
Using the same reference branch and conventions as
Appendix~\ref{app:prad_pgas},
\begin{equation}
\frac{H}{r}=\frac{3\kappa_{\rm es}m_p}{2\sigma_T}
\frac{\lambda_{\rm E}\mathcal J}{\eta_{\rm rad}(r/r_g)}
\simeq0.03
\left(\frac{\lambda_{\rm E}}{0.1}\right)
\left(\frac{0.1}{\eta_{\rm rad}}\right)
\left(\frac{10r_g}{r}\right)
\frac{\mathcal J}{0.225}
\end{equation}
With $\Sigma=\dot M\mathcal J/(3\pi\alpha\Omega_KH^2)$,
$r=10r_g$, $r_{\rm in}=6r_g$, and $\alpha=\eta_{\rm rad}=0.1$,
$\lambda_{\rm E}=0.01$--0.1 gives $H/r\simeq0.0029$--0.029
and $\Sigma\simeq2.9\times10^4$--$2.9\times10^3\,{\rm g\,cm^{-2}}$,
independent of $M_\bullet$ on this branch. For a one-zone optical depth $\tau\simeq\kappa_{\rm es}\Sigma$, the diffusion
and steady inflow times obey
\begin{equation}
\frac{t_{\rm diff}}{t_{\rm in}}
\sim\frac{\tau H/c}{2r^2\mathcal J/(3\nu)}
\simeq4.9\times10^{-3}
\left(\frac{\lambda_{\rm E}}{0.1}\right)^2
\quad(r=10r_g,\ \eta_{\rm rad}=0.1).
\end{equation}
Thus photons escape before inflow over this range. This tests photon trapping, not thermal stability or evaporation. At $\lambda_{\rm E}\sim1$,
$H/r\sim0.3$ and the thin-disk approximation becomes marginal.

\section{Location of the self-gravity boundary}
\label{app:toomre}

For a Keplerian disk $Q=c_{\rm eff}\Omega_K/(\pi G\Sigma)$ with $\dot M\mathcal{J}=3\pi\alpha c_{\rm eff}H\Sigma$ and $H=c_{\rm eff}/\Omega_K$, we get $Q=3\alpha c_{\rm eff}^3/(G\dot M\mathcal{J})$. On the inner radiation branch it gives $Q\gtrsim4\times10^5$ at $10r_g$ and $\gtrsim7\times10^2$ at $100r_g$ over $M_\bullet=10^6$--$10^8M_\odot$ and $\lambda_{\rm E}=0.01$--1, so the inner disk is not likely self-gravitating. The standard disk reaches $Q\simeq1$ only at $r_{\rm sg}\sim10^3$--a few$\times10^5\,r_g$ ($\sim0.005$--$0.1\,{\rm pc}$), where opacity, irradiation, magnetic support, and star formation set the precise boundary \citep{goodman03,chen+23}.

\section{Comptonization in a collisionless plasma}
\label{app:thermal_compton}

Here ``collisionless'' means Coulomb collisions do not control reconnection, turbulent dissipation, ion acceleration, or electron--ion exchange: at $R\sim10r_g$, $\tau_T\sim1$, the Coulomb mean free path far exceeds the skin depth, while ion relaxation and electron--ion exchange times exceed $R/c$. Electron--electron scattering may still relax a weakly collisional thermal core, and denser transition layers can be more collisional \citep{bambic+24a}.

Comptonization of UV seed photons by quasi-thermal electrons can produce an X-ray component even in a collisionless plasma, since collective plasma processes can isotropize particle momenta without requiring Coulomb collisions \citep[e.g.,][]{kunz+14b,sironi+15b}. For seed photons with $E_0\sim5$--$20\,{\rm eV}$, the mean energy amplification per scattering in the Thomson regime is $A\simeq1+4\theta_e+16\theta_e^2$, where $\theta_e=kT_e/(m_ec^2)$ \citep{rybicki+79,sunyaev+80}. Repeated scattering by electrons with $kT_e\sim50\,{\rm keV}$ can build an X-ray tail with a cutoff of order a few $kT_e$ \citep{sunyaev+80,haardt+91}.


\paragraph*{Turbulent heating and photon escape.}
If turbulent magnetic energy $U_B$ dissipates on an eddy time $t_0$, writing the total photon amplification as $\mathcal{A}_{\rm C}$, steady balance of the photon energy-gain rate against heating gives $U_{\rm seed}(\mathcal{A}_{\rm C}-1)/t_{\gamma,{\rm esc}}\sim U_B/t_0$, so $U_X\simeq \mathcal{A}_{\rm C} U_{\rm seed}\simeq U_{\rm seed}+U_B(t_{\gamma,{\rm esc}}/t_0)$. When the Comptonized component dominates, we get $U_X\sim U_B(t_{\gamma,{\rm esc}}/t_0)$, i.e. $\kappa_{\rm x}=  U_B/U_X\simeq t_0/t_{\gamma,{\rm esc}}$. The X-ray power is set by turbulent heating, while raising $U_{\rm seed}$ mainly lowers the amplification $\mathcal{A}_{\rm C}$ and softens the spectrum.

\paragraph*{Inverse-Compton cooling length.}
For an electron with Lorentz factor $\gamma$ in an isotropic radiation field of energy density $U_{\rm rad}$, the Thomson-regime IC cooling rate is $|d\gamma/dt|_{\rm IC}=(4/3)(\sigma_T c/m_e c^2)U_{\rm rad}\gamma^2$ \citep{rybicki+79}. For a region of size $R$ with soft-photon luminosity $L$ and compactness $\ell= L\sigma_T/(R m_e c^3)$ \citep{lightman+87}:
\begin{equation}\label{eq:app_psiIC}
	\Psi_{\rm IC} = \frac{\lambda_{\rm cool}}{R}=\frac{3\pi}{\ell\gamma}.
\end{equation}
In a compact UV-bright region (large $\ell$), IC losses occur over a small fraction of $R$. In the Thomson limit, upscattering UV photons of energy $E_0$ to X-ray energy $E_X$ requires $\gamma_{\min}\sim(E_X/E_0)^{1/2}$, so $\lambda_{\rm cool}/R\sim 3\pi/(\ell\gamma_{\min})$. For $E_0\sim 10\,{\rm eV}$, $E_X\sim 1\,{\rm keV}$, $\gamma_{\min}\sim 10$, and $\ell\sim 100$: $\lambda_{\rm cool}/R\sim 10^{-2}$, so electrons cool well within the source.

\paragraph*{Optical depth and the Compton $y$-parameter.}
The Thomson optical depth $\tau_T= n_e\sigma_T R$ and Compton parameter $y\sim 4\theta_e\max(\tau_T,\tau_T^2)$ \citep{rybicki+79,sunyaev+80} control Comptonization efficiency. For unsaturated thermal Comptonization with $y\sim 1$, in the optically thick regime: $\tau_T\sim(4\theta_e)^{-1/2}$. For typical coronal temperatures $kT_e\sim 30$--$100\,{\rm keV}$ ($\theta_e\sim 0.06$--$0.2$), this yields $\tau_T\sim 1$--$2$, consistent with inference from AGN spectra \citep{haardt+91,fabian+15}.

\paragraph*{Combined condition.}
A compact UV-bright region with $\ell\gtrsim10$ can cool electrons efficiently by IC, but producing a strong thermal-Compton X-ray component also requires sufficient scattering depth. At $\tau_T\ll0.1$, most UV photons escape without scattering; at $\tau_T\gg5$, repeated scattering at typical coronal temperatures tends toward saturated Comptonization and a Wien-like bump. An optical depth $\tau_T\sim\mathcal{O}(1)$ therefore favors the unsaturated thermal-Compton spectrum observed in AGN \citep{sunyaev+80,haardt+91}.

\bsp	
\label{lastpage}
\end{document}